  \documentclass[aps,prd,reprint,superscriptaddress,nofootinbib,floatfix,longbibliography]{revtex4-2}

  \usepackage[T1]{fontenc}
  \usepackage[utf8]{inputenc}
  \usepackage[english]{babel}

  \usepackage[dvipsnames]{xcolor}

  \usepackage{amsmath,amsbsy,latexsym,cancel}

  \usepackage{newtxtext}
 
  \usepackage{amssymb,amsfonts}
  \usepackage{bm} 
  \usepackage{mathrsfs}
  \usepackage{euscript}

  \usepackage{array}
  \usepackage{dcolumn}
  \usepackage{longtable}
  \usepackage{rotating} 
  \usepackage{booktabs} 
  \usepackage[caption=false]{subfig}
  \usepackage{placeins}

  \usepackage{multirow}
  \usepackage{enumerate} 

  \usepackage{silence}
  \usepackage{orcidlink}
  \usepackage{hyperref}
  \hypersetup{
    colorlinks=true,
    linkcolor=blue,
    citecolor=blue,
    urlcolor=blue
  }
\begin{document}

   \renewcommand{\sectionautorefname}{Section}
  \renewcommand{\subsectionautorefname}{Section}
  \renewcommand{\subsubsectionautorefname}{Section}

   \title{Non-Standard Interactions and Light \texorpdfstring{$Z'$}{Z'} Bosons from \texorpdfstring{CE$\nu \mathcal{N}$S}{CEvNS} Data at \texorpdfstring{\textsc{Conus}$+$}{CONUS+}}

  \author{Ch.~M.~Benavides~\orcidlink{0009-0006-9765-1169}}
  \email{chmbenavides@gmail.com}
  \affiliation{Departamento de Física, Universidad de Nariño, Pasto, Colombia}

  \author{C.~S.~Muñoz-López~\orcidlink{0009-0004-0189-3266}}
  \email{cristiansantiagocsml123@gmail.com}
  \affiliation{Departamento de Física, Universidad de Nariño, Pasto, Colombia}

  \author{B.C.~Cañas~\orcidlink{0000-0002-7334-5304}}
  \email{blanca.canas@unipamplona.edu.co}
  \affiliation{Departamento de Física, Universidad de Pamplona, Pamplona, Colombia}

  \author{E.~Rojas~\orcidlink{0000-0002-7412-1345}}
  \email{rojas@udenar.edu.co}
  \affiliation{Departamento de Física, Universidad de Nariño, Pasto, Colombia}

  \date{\today}

  \begin{abstract}

    We analyze the first  \textsc{Conus}$+$   reactor-antineutrino spectrum to constrain nonstandard neutrino interactions (NSI) with quarks and light vector $Z'$ mediators in $E_6$ and leptophilic $U(1)_{L_a-L_b}$ models. The prediction includes coherent elastic antineutrino--nucleus  scattering (CE  $\nu\mathcal N$S) and elastic antineutrino--electron scattering (E$\nu e$S), with  Standard Model (SM)--mediator interference in each channel and an analytically profiled normalization uncertainty. We formulate $L_e-L_\mu$ and $L_e-L_\tau$ in a common charge convention, combining direct electron scattering with the nuclear interaction induced by photon-- $Z'$   mixing. The data remain compatible with the SM. Electronic scattering has little effect on the quark-NSI and fixed-coupling $E_6$ angular regions, but substantially strengthens the low-mass coupling limits of the $E_6$ benchmarks. The leptophilic models share a common low-mass plateau dominated by electronic scattering, while their limits separate at larger mediator masses through their different nuclear loop contributions.

   \end{abstract}
  
  \maketitle

\section{Introduction}
\label{sec:introduction}

  Coherent elastic neutrino--nucleus scattering (CE$\nu \mathcal{N}$S) is a Standard Model (SM) process in which a neutrino interacts coherently and elastically with an atomic nucleus. This process was theoretically predicted by Daniel Z. Freedman in 1974~\cite{Freedman:1973yd} and was discovered in 2017 by the COHERENT collaboration using a CsI[Na] detector~\cite{COHERENT:2017ipa}. Subsequent new observations of CE$\nu \mathcal{N}$S have been reported by COHERENT using a liquid argon detector in 2021~\cite{COHERENT:2020iec} and a germanium detector in 2025~\cite{COHERENT:2024axu}. All these measurements used neutrinos produced at the Spallation Neutron Source (SNS). In parallel, the PandaX and XENON-nT dark-matter experiments reported indications of solar CE$\nu \mathcal{N}$S in liquid xenon in 2024~\cite{PandaX:2024muv,XENON:2024ijk}. Recently, the \textsc{Conus}$+$ collaboration reported the first direct observation of CE$\nu \mathcal{N}$S with a reactor antineutrino source, with a statistical significance of  3.7$\sigma$ ~\cite{Ackermann:2025obx}. 

  Based on the results reported by the \textsc{Conus}$+$ collaboration, several phenomenological studies have been carried out to evaluate the sensitivity of CE$\nu \mathcal{N}$S not only as a precision electroweak test, but also to various low-momentum-transfer new-physics scenarios, including electromagnetic neutrino properties, sterile-neutrino production and nonstandard interactions (NSI)~\cite{AtzoriCorona:2025ygn,AtzoriCorona:2025xgj,DeRomeri:2025csu,Chattaraj:2025fvx,Alpizar-Venegas:2025wor,CONUS:2026uhz}. In particular, constraints on NSI using \textsc{Conus}$+$ data have already been studied in recent analyses. Ref.~\cite{DeRomeri:2025csu} provides the most directly comparable spectral treatment and is used here as the main validation benchmark for our numerical implementation, while Refs.~\cite{Alpizar-Venegas:2025wor,CONUS:2026uhz} further illustrate the sensitivity of \textsc{Conus}/\textsc{Conus}$+$ data to NSI and related   beyond-the-Standard-Model (BSM)  scenarios. The NSI sector is therefore included in the present work as an independent consistency check, as a self-contained baseline within the same statistical pipeline, and as a reference framework for the comparison with light vector $Z'$ scenarios. We also discuss the associated two-dimensional degeneracy structure and its one-dimensional projections.

  Our main focus is the   sensitivity  of the \textsc{Conus}$+$   excess spectrum to  light vector bosons   with specified charge assignments. We consider  $E_6$   benchmarks and  the anomaly-free leptophilic symmetries   $U(1)_{L_a-L_b}$, where $a,b\in\{e,\mu,\tau\}$ and $a\ne b$~\cite{He:1991qd}. For $L_e-L_\mu$ and $L_e-L_\tau$, the mediator couples directly to electrons and reactor antineutrinos, while its nuclear interaction arises through charged-lepton loops that generate photon-- $Z'$   kinetic mixing~\cite{Cadeddu:2020nbr,AtzoriCorona:2022moj}. A consistent prediction therefore includes both direct elastic antineutrino--electron scattering (E$\nu e$S) and the  loop-induced   nuclear channel.

  Electronic and nuclear recoils contribute to the measured ionization spectrum. We add their predicted rates in each reconstructed-energy bin, retaining the SM contribution and mediator interference within each channel. The small SM electronic rate does not justify neglecting a mediator-enhanced electronic signal. This distinction is especially relevant for leptophilic models, whose direct E$\nu e$S cross sections are identical for $L_e-L_\mu$ and $L_e-L_\tau$ \cite{Demirci:2026nju}, whereas their nuclear responses depend on different internal lepton masses. We retain the flavor charge explicitly throughout the mixing calculation to fix the normalization and relative sign of the SM and mediator amplitudes .

    We use  a common binned-$\chi^2$   analysis with a profiled normalization to study quark NSI,  $E_6$   charge assignments, and both leptophilic models. Generic mediator couplings provide the common formalism; numerical constraints are obtained for the specified models. The combined prediction leaves the quark-NSI and fixed-coupling  $E_6$   angular results practically unchanged at the evaluated resolution, while electronic scattering controls the strongest low-mass mediator limits. The nuclear contribution remains relevant at larger masses and distinguishes the two leptophilic models. The quantitative comparisons and plateau definitions are given in \autoref{sec:results} and \autoref{sec:profiled-analysis}, respectively.

  The paper is organized as follows .  \autoref{sec:formalism} defines the interactions and cross sections, \autoref{sec:analysis} describes the signal prediction and statistical treatment, \autoref{sec:results} presents the constraints, and \autoref{sec:conclusions} summarizes their implications.

\section{Theoretical Formalism}
\label{sec:formalism}

   In this section, we present the   theoretical description of CE$\nu \mathcal{N}$S and E$\nu e$S. We first give the SM cross sections and then introduce NSI with quarks . Next, we   consider  light vector $Z'$ mediators,   starting from generic couplings to neutrinos, electrons, and quarks, and specialize them to $E_6$ and leptophilic $U(1)_{L_a-L_b}$ models. This formulation specifies the nuclear and electronic contributions within the same coupling conventions.

\subsection{  Standard Model  cross   sections }
 \label{sec:sm-cross-sections}

     CE$\nu \mathcal{N}$S is a SM neutral-current process

    \begin{equation}
      \bar\nu_f + {}^{A}_{Z}\mathcal{N} \;\rightarrow\; \bar\nu_f + {}^{A}_{Z}\mathcal{N},
      \qquad
      f = e,\mu,\tau.
      \label{eq:cevns-process}
    \end{equation} 

     For the reactor source, $f=e$.  At the fundamental level,   antineutrinos  interact with quarks through the exchange of a $Z^{0}$ boson; however, in the coherent regime, the momentum transfer is small enough that the internal structure of the nucleus is not resolved. Here, the momentum transferred in the interaction is smaller than the inverse of the size of the nucleus; as a result, the scattering amplitudes of the nucleons add up coherently~\cite{Freedman:1973yd}. The dominant contribution is the vector nuclear current, while axial contributions are not coherently enhanced and are neglected in the present analysis.

    The SM prediction for the differential cross section of CE$\nu \mathcal{N}$S as a function of the nuclear recoil energy  $T_{\mathcal{N}}$ reads\footnote{In this work we employ natural units so that $\hbar = c = 1$.} \cite{Freedman:1973yd,DeRomeri:2025csu}

    \begin{equation}
    \begin{aligned}
      \left. \frac{d\sigma_{\bar\nu_e\mathcal{N}}}{dT_{\mathcal{N}}} \right|_{\text{SM}} = &
      \frac{G_F^2\,m_{\mathcal{N}}}{\pi}
      \left(Q_{V}^{\text{SM}}\right)^2
      F_W^2(|\mathbf{q}|^2)\\
       & \times
      \left[
      1 - \frac{m_{\mathcal{N}} T_{\mathcal{N}}}{2E_{\bar\nu_e}^2}
      - \frac{T_{\mathcal{N}}}{E_{\bar\nu_e}}
      \right],        
    \end{aligned}
    \label{eq:cross-section-sm}
    \end{equation} 
    where $G_F$ is the Fermi constant, $m_{\mathcal{N}}$ is the target nuclear mass, $E_{\bar\nu_e}$ is the incident electron-antineutrino energy and $|\mathbf{q}|\simeq\sqrt{2m_{\mathcal{N}}T_{\mathcal{N}}}$ is the magnitude of the momentum transfer. The tree-level weak nuclear charge is given by
  
    \begin{equation}
      Q_{V}^{\text{SM}} = g_{V}^p\,Z + g_V^n\,N,
      \label{eq:sm-nuclear-charge}
    \end{equation} 
      where $Z$ and $N$ count protons and neutrons , respectively. The neutrino--nucleon vector couplings are 
      \begin{equation}
      g_{V}^p = \frac{1-4\sin^2\theta_W}{2},
      \qquad
      g_V^n = -\frac{1}{2},
      \label{eq:sm-nucleon-couplings}
    \end{equation} 
    where $\theta_W$ is the weak mixing angle, with $\sin^2\theta_W=0.23857(5)$, for the low-energy regime relevant to CE$\nu \mathcal{N}$S~\cite{ParticleDataGroup:2024cfk}. The weak nuclear form factor $F_W(|\mathbf{q}|^2)$ encodes the spatial distribution of the weak charge inside the nucleus and describes the progressive loss of coherence as the momentum transfer $|\mathbf{q}|$ increases. In the low-energy regime relevant for reactor antineutrinos (  $E_{\bar\nu_e} \lesssim 10$  MeV), the typical momentum transfer satisfies: $|\mathbf{q}| R_A \ll 1$, which implies $F_W(|\mathbf{q}|^2) \simeq 1$.

     Reactor antineutrinos also scatter elastically from electrons through $\bar\nu_e+e^-\to\bar\nu_e+e^-$. In the SM, this process receives a neutral-current contribution from $Z$ exchange and a charged-current contribution from $W$ exchange. For an unpolarized free electron initially at rest, the differential cross section is~\cite{DeRomeri:2025csu}
    \begin{equation}
      \begin{aligned}
        \left.\frac{d\sigma_{\bar\nu_e e}}{dT_e}\right|_{\mathrm{SM}}
        ={}&\frac{G_F^2m_e}{2\pi}
        \bigg[
          (c_V-c_A)^2\\
          &+(c_V+c_A)^2
          \left(1-\frac{T_e}{E_{\bar\nu_e}}\right)^2\\
          &-(c_V^2-c_A^2)\frac{m_eT_e}{E_{\bar\nu_e}^2}
        \bigg],
      \end{aligned}
      \label{eq:cross-section-sm-eves}
    \end{equation}
    where $T_e$ is the electron recoil kinetic energy and $m_e$ is the electron mass. In the convention where the electron current contains $\gamma^\mu(c_V-c_A\gamma^5)$, the coefficients are
    \begin{equation}
      c_V=\frac{1}{2}+2\sin^2\theta_W,
      \qquad
      c_A=\frac{1}{2}.
      \label{eq:sm-eves-couplings}
    \end{equation}
    These coefficients include both charged and neutral currents and their interference. The placement of the factor $(1-T_e/E_{\bar\nu_e})^2$ in Eq.~\eqref{eq:cross-section-sm-eves} corresponds specifically to antineutrinos. This is a cross section per free electron; atomic binding and the number of accessible electrons in germanium enter separately when constructing an electronic event rate.

 \subsection{Neutrino Non-Standard Interactions}
\label{sec:nsi}
    NSI provides a model-independent parametrization of new-physics effects at low energies, encoded in effective four-fermion operators~\cite{Ohlsson:2012kf, Miranda:2015dra, Farzan:2017xzy}. Since CE$\nu \mathcal{N}$S is a neutral-current process~\cite{Freedman:1973yd,Giunti:2019xpr,Canas:2019fjw}, only neutral-current NSI are considered. The corresponding effective Lagrangian is given by~\cite{Papoulias:2015vxa}

    \begin{equation}
    \resizebox{0.43\textwidth}{!}{$
      \mathcal{L}_{\text{NC}}^{\text{NSI}} =
      -2\sqrt{2} G_F
      \sum_{f,f'} 
      \varepsilon_{ff'}^{fC}
      \left( \bar{\nu}_f \gamma^\mu P_L \nu_{f'} \right)
      \left( \bar{\psi} \gamma_\mu P_C \psi \right),
    $}  
      \label{eq:nsi-lagrangian}
    \end{equation} 
    where $f,f' = e,\mu,\tau$ denote flavor indices, $C = L,R$ indicates chirality, and $\psi$ represents SM fermions. The parameters $\varepsilon_{ff'}^{fC}$ quantify the strength of the NSI relative to the Fermi interaction. These include both non-universal ($f=f'$) and  flavor-changing ($f \neq f'$) contributions. Since axial contributions are suppressed in CE$\nu \mathcal{N}$S~\cite{DeRomeri:2025csu, Barranco:2005yy}, only vector interactions are retained, defined as
      \begin{equation}
      \varepsilon_{ff'}^{fV} =
      \varepsilon_{ff'}^{fL} +
      \varepsilon_{ff'}^{fR}.
      \label{eq:nsi-vector-couplings}
    \end{equation}

    In this framework, the CE$\nu \mathcal{N}$S differential cross section in Eq.~\eqref{eq:cross-section-sm} is modified through a redefinition of the weak charge at the amplitude level. The squared effective charge, including NSI contributions, is given by~\cite{Canas:2019fjw, Barranco:2005yy, DeRomeri:2025csu}

    \begin{equation}
    \resizebox{0.43\textwidth}{!}{%
\mbox{
$
    \begin{aligned}
      \left(Q_V^{\text{NSI}}\right)^2 =&
      \left|
      Z \left( g_V^p + 2\varepsilon_{ff}^{uV} + \varepsilon_{ff}^{dV} \right)
      +
      N \left( g_V^n + \varepsilon_{ff}^{uV} + 2\varepsilon_{ff}^{dV} \right)
      \right|^2
      \\
      &+
      \sum_{f \neq f'}
      \left|
      Z \left( 2\varepsilon_{ff'}^{uV} + \varepsilon_{ff'}^{dV} \right)
      +
      N \left( \varepsilon_{ff'}^{uV} + 2\varepsilon_{ff'}^{dV} \right)
      \right|^2.
    \end{aligned}
    $
}
}
      \label{eq:nsi-nuclear-charge}
    \end{equation} 

    This expression explicitly shows that diagonal NSI ($f=f'$) interfere coherently with the SM contribution, while off-diagonal terms ($f \neq f'$) contribute incoherently at the level of the squared amplitude. In the analysis of \textsc{Conus}$+$ reactor antineutrino data, the incoming   antineutrino  flavor is $f=e$. Consequently, the relevant NSI parameters are $\varepsilon_{ef'}^{uV}$ and $\varepsilon_{ef'}^{dV}$, where $f'=e,\mu,\tau$.
     Only NSI with quarks are included here. NSI with electrons can also be defined, but involve independent coefficients that are outside the scope of the present parametrization; the quark operators above do not modify the E$\nu e$S process at tree level.

\subsection{Light vector \texorpdfstring{$Z'$}{Z'} mediators}
\label{sec:light-vector-mediators}

    In this subsection, we consider a generic vector mediator   coupled to neutrinos, electrons, and quarks, followed by two classes of charge assignments. The first  consists of $E_6$-motivated   models  in the  two-angle parametrization, as shown in the  Sanson--Flamsteed   (SF) projection ~\cite{Langacker:2008yv,Rojas:2015tqa}. The   second consists of leptophilic $U(1)_{L_a-L_b}$ models, whose  coupling to nuclei is   induced through $\gamma$ --$Z'$ kinetic mixing.   We give common cross sections for direct $Z'$ exchange and then specify the couplings for each model and the additional loop-induced nuclear contribution for the leptophilic models.

\subsubsection{Generic vector mediator}

\label{sec:generic-vector-mediator}
       A new neutral vector boson $Z'$ with mass $m_{Z'}$ is considered. Its interaction Lagrangian with SM fermions is written as~\cite{Langacker:2008yv}
        \begin{equation}
      \resizebox{0.43\textwidth}{!}{$
        \mathcal{L}_{Z'}^{\text{int}} = -\frac{g_{Z'}}{2} \sum_{\psi} \bar{\psi}\,\gamma^{\mu} \bigl[g_{V}^{Z'}(\psi) - g_{A}^{Z'}(\psi)\gamma^{5}\bigr]\psi\,Z'_{\mu},
      $}
      \label{eq:zprime-interaction}
      \end{equation}

       where $g_{Z'}$ is the coupling constant of the new $U(1)'$ gauge group. The vector and axial-vector coefficients are expressed in terms of the left- and right-handed chiral charges $Q^{Z'}_{L, R}(\psi)$~\cite{Langacker:2017uah, Rojas:2015tqa}.
      \begin{equation}
      \begin{aligned}
        g_V^{Z'}(\psi) =& Q_L^{Z'}(\psi) + Q_R^{Z'}(\psi), \\
        g_A^{Z'}(\psi) =& Q_L^{Z'}(\psi) - Q_R^{Z'}(\psi).
      \end{aligned}
      \label{eq:vector-couplings}
      \end{equation}

       For the left-handed neutrino fields, the couplings satisfy $g_V^{Z'}\boldsymbol{(}\bar\nu_f\boldsymbol{)}=g_A^{Z'}\boldsymbol{(}\bar\nu_f\boldsymbol{)}=Q_L^{Z'}\boldsymbol{(}\bar\nu_f\boldsymbol{)}$, with $f=e,\mu,\tau$. The argument $\bar\nu_f$ labels the incident reactor channel; these coefficients refer to the charge of the neutrino field in the Lagrangian, with no additional sign reversal for an external antineutrino. Throughout this subsection, the couplings are real and the antineutrino mass is neglected.

       The coherent nuclear matrix element of the $Z'$ vector current can be written in terms of the effective coherent nuclear vector charge  ~\cite{Papoulias:2017qdn}:
      \begin{equation}
      \begin{aligned}
        Q_{\mathcal{N},V}^{Z'} = &  Z\bigl[2g_V^{Z'}(u) + g_V^{Z'}(d)\bigr]\\
          &+ N\bigl[g_V^{Z'}(u) + 2g_V^{Z'}(d)\bigr].          
      \end{aligned}
      \label{eq:zprime-nuclear-charge}
      \end{equation} 
    For an incident $\bar\nu_f$, the  exchange of $Z'$ then modifies the CE$\nu \mathcal{N}$S cross section in Eq.~\eqref{eq:cross-section-sm} as a rescaling of the SM prediction,
        \begin{equation}
        \left.\frac{d\sigma_{\bar\nu_f\mathcal{N}}}{dT_{\mathcal{N}}}\right|_{Z'+\text{SM}} = \mathcal{G}_{Z'}^2\, \left.\frac{d\sigma_{\bar\nu_f\mathcal{N}}}{dT_{\mathcal{N}}}\right|_{\text{SM}},
        \label{eq:cross-section-zprime-cevns}
      \end{equation}

       with the scaling factor~\cite{Papoulias:2017qdn}

      \begin{equation}
      \resizebox{0.42\textwidth}{!}{$
        \mathcal{G}_{Z'}^2 = \left|1 + \frac{1}{2\sqrt{2}\,G_F} \frac{Q_{\mathcal{N},V}^{Z'}}{Q_V^{\text{SM}}} \frac{g_{Z'}^2\,g_V^{Z'}\boldsymbol{(}\bar\nu_f\boldsymbol{)}}{2m_{\mathcal{N}}T_{\mathcal{N}} + m_{Z'}^2}\right|^2.
      $}
      \label{eq:zprime-cevns-rescaling}
      \end{equation} 

       For reactor E$\nu e$S, the SM+$Z'$ differential cross section per free electron is given by~\cite{Dent:2016wcr, Demirci:2026nju}
      \begin{equation}
        \begin{aligned}
          \left.\frac{d\sigma_{\bar\nu_e e}}{dT_e}\right|_{\mathrm{SM}+Z'}
          ={}&\frac{G_F^2m_e}{2\pi}
          \bigg[
            (c_V^{\mathrm{eff}}-c_A^{\mathrm{eff}})^2\\
            &+(c_V^{\mathrm{eff}}+c_A^{\mathrm{eff}})^2
            \left(1-\frac{T_e}{E_{\bar\nu_e}}\right)^2\\
            &-\bigl[(c_V^{\mathrm{eff}})^2-(c_A^{\mathrm{eff}})^2\bigr]
            \frac{m_eT_e}{E_{\bar\nu_e}^2}
          \bigg],
        \end{aligned}
        \label{eq:cross-section-zprime-eves}
      \end{equation}
      where the effective couplings, evaluated at $T_e$, are defined using the normalization of Eq.~\eqref{eq:zprime-interaction} by
      \begin{equation}
        \begin{aligned}
          c_{V,A}^{\mathrm{eff}}\boldsymbol{(}T_e\boldsymbol{)}
          &=
          c_{V,A}
          +\eta_{Z'}\boldsymbol{(}T_e\boldsymbol{)}
          g_{V,A}^{Z'}\boldsymbol{(}e\boldsymbol{)},\\
          \eta_{Z'}\boldsymbol{(}T_e\boldsymbol{)}
          &=
          \frac{g_{Z'}^2g_V^{Z'}\boldsymbol{(}\bar\nu_e\boldsymbol{)}}{2\sqrt2G_F(m_{Z'}^2+2m_eT_e)}.
        \end{aligned}
        \label{eq:eves-effective-couplings}
      \end{equation}
      Here $c_V$ and $c_A$ are given in Eq.~\eqref{eq:sm-eves-couplings}. Equation~\eqref{eq:cross-section-zprime-eves} includes SM--$Z'$ interference and the squared $Z'$ amplitude contribution, and reduces to Eq.~\eqref{eq:cross-section-sm-eves} as $g_{Z'}\to0$.

 \FloatBarrier
\subsubsection{\texorpdfstring{$E_6$}{E6}-motivated \texorpdfstring{$Z'$}{Z'} models}

\label{sec:e6-models}

       Grand-unified extensions based on $E_6$ naturally predict additional anomaly-free $U(1)'$ gauge factors and therefore neutral gauge bosons $Z'$~\cite{Langacker:2008yv}. Since $E_6$ has rank six, the physically relevant Abelian directions beyond the SM can be organized in terms of two angular parameters. The usual basis is associated with the breaking chains $E_6\to SO(10)\times U(1)_\psi$ and $SO(10)\to SU(5)\times U(1)_\chi$, which define the $Z_\psi$ and $Z_\chi$ directions, respectively. A third component, proportional to the SM hypercharge direction $Z_Y$, can appear when $U(1)_Y$--$U(1)'$ kinetic mixing is allowed~\cite{Langacker:2008yv,Rojas:2015tqa}.

      In the two-angle parameterization used below, $\beta$ controls the admixture of the $U(1)_\chi$ and $U(1)_\psi$ directions, while $\alpha$ parametrizes the hypercharge admixture induced by kinetic mixing. The corresponding angular domain, $\alpha,\beta\in[-\pi/2,\pi/2]$ , is displayed in this work using the Sanson--Flamsteed projection \cite{Rojas:2015tqa}, so that each benchmark model is represented by a specific direction in the $E_6$ angular space.   For the fixed-coupling angular scan, we use the unification-motivated normalization
      \begin{equation}
        g_{Z'} = \sqrt{\frac{5}{3}}\,g\tan\theta_W \simeq 0.46,
        \label{eq:normalization_Langacker}
      \end{equation} 
      in agreement with Eqs.~(44) and (49) of Ref.~\cite{Langacker:2008yv}  ; $g$ is the SM $SU(2)_L$ gauge coupling. In the benchmark mass--coupling scans, $g_{Z'}$ is varied independently.  The fermion charges in the fundamental $\mathbf{27}$ representation of $E_6$ are listed in   \autoref{tab:e6-charges} .

      In this angular basis, the physical $Z'$ is written as a linear combination of three gauge eigenstates~\cite{Rojas:2015tqa},

      \begin{equation}
      \begin{aligned}
        Z' =& \cos\alpha\cos\beta\,Z_{\chi} + \sin\alpha\cos\beta\,Z_{Y}  \\
        &+ \sin\beta\,Z_{\psi},          
      \end{aligned}
        \label{eq:e6-gauge-basis}
      \end{equation} 
 
    \begin{table}[ht]
        \centering
        \caption{\label{tab:e6-charges}Fermion charges in the $E_6$ fundamental representation.~\cite{Langacker:2008yv}}
        \begin{tabular}{lccc}
          \toprule \toprule
          Field & $2\sqrt{10}\,Q^{\chi}_{E_6}$ & $2\sqrt{6}\,Q^{\psi}_{E_6}$ & $\sqrt{5/3}\,Q_Y^{E_6}$ \\
          \midrule
          $\nu_L$ & $3$  & $1$  & $-1/2$ \\
          $e_L$   & $3$  & $1$  & $-1/2$ \\
          $e_R$   & $1$  & $-1$ & $-1$   \\
          $u_L$   & $-1$ & $1$  & $1/6$  \\
          $u_R$   & $1$  & $-1$ & $2/3$  \\
          $d_L$   & $-1$ & $1$  & $1/6$  \\
          $d_R$   & $-3$ & $-1$ & $-1/3$ \\
          \bottomrule \bottomrule
        \end{tabular}
      \end{table}

       which induces the chiral charges

       \begin{widetext}
        \begin{equation}
        Q_{L,R}^{Z'}(\psi) =
        \cos\alpha\cos\beta\,Q^{\chi}_{E_6}(\psi_{L,R}) +
        \sin\alpha\cos\beta\,Q^{Y}_{E_6}(\psi_{L,R}) +
        \sin\beta\,Q^{\psi}_{E_6}(\psi_{L,R}).
        \label{eq:e6-chiral-charges}
      \end{equation} 
      \end{widetext}    
      The   couplings entering the nuclear and electronic cross sections are obtained  from Eq.~\eqref{eq:vector-couplings}   and \autoref{tab:e6-charges}:
      \begin{equation}
        \begin{aligned}
          g_V^{Z'}\boldsymbol{(}\bar\nu_f\boldsymbol{)}
          &=
          \frac{3}{2\sqrt{10}}\cos\alpha\cos\beta
          -\frac{1}{2}\sqrt{\frac{3}{5}}\sin\alpha\cos\beta\\
          &\quad+\frac{1}{2\sqrt6}\sin\beta,\\
          g_V^{Z'}\boldsymbol{(}u\boldsymbol{)}
          &=\frac{5}{6}\sqrt{\frac{3}{5}}\sin\alpha\cos\beta,\\
          g_V^{Z'}\boldsymbol{(}d\boldsymbol{)}
          &=-\frac{2}{\sqrt{10}}\cos\alpha\cos\beta
          -\frac{1}{6}\sqrt{\frac{3}{5}}\sin\alpha\cos\beta,
        \end{aligned}
        \label{eq:e6-neutrino-quark-couplings}
      \end{equation}
      with $g_A^{Z'}\boldsymbol{(}\bar\nu_f\boldsymbol{)}=g_V^{Z'}\boldsymbol{(}\bar\nu_f\boldsymbol{)}$. The electron couplings are
      \begin{equation}
        \begin{aligned}
          g_V^{Z'}\boldsymbol{(}e\boldsymbol{)}
          &=\frac{2}{\sqrt{10}}\cos\alpha\cos\beta
          -\frac{3}{2}\sqrt{\frac{3}{5}}\sin\alpha\cos\beta,\\
          g_A^{Z'}\boldsymbol{(}e\boldsymbol{)}
          &=\frac{1}{\sqrt{10}}\cos\alpha\cos\beta
          +\frac{1}{2}\sqrt{\frac{3}{5}}\sin\alpha\cos\beta\\
          &\quad+\frac{1}{\sqrt6}\sin\beta.
        \end{aligned}
        \label{eq:e6-electron-couplings}
      \end{equation}
      Equations~\eqref{eq:e6-neutrino-quark-couplings} and \eqref{eq:e6-electron-couplings} specify the couplings for both channels: the quark couplings enter Eq.~\eqref{eq:zprime-nuclear-charge}, whereas the electron couplings enter Eq.~\eqref{eq:eves-effective-couplings}. The corresponding cross sections are then obtained from Eqs.~\eqref{eq:cross-section-zprime-cevns} and \eqref{eq:cross-section-zprime-eves}, without further changes to their kinematic dependence.

        The nuclear mediator amplitude is  controlled by the product   $g_V^{Z'}\boldsymbol{(}\bar\nu_f\boldsymbol{)}Q_{\mathcal{N},V}^{Z'}$; the electronic mediator amplitude instead depends on the products of the neutrino charge with the electron vector and axial couplings. Consequently, changing the angular direction modifies both components of the predicted spectrum at fixed  $m_{Z'}$ and $g_{Z'}$. Special limiting cases of phenomenological interest, including $B\!-\!L$, $\chi$, $\psi$, and the down-, neutron-, and proton-phobic directions, are obtained for particular values of $(\alpha,\beta)$ within this $E_6$ angular space. These benchmarks are representative directions rather than arbitrary examples, and are summarized in   \autoref{tab:e6-model-charges}. \cite{Rojas:2015tqa}

      \begin{table}[ht]
        \caption{\label{tab:e6-model-charges}Charges associated with different $U(1)'$ groups in the two-angle $E_6$ basis.~\cite{Rojas:2015tqa}}
        \centering
        \begin{tabular}{l c c c}
          \toprule \toprule
          $U(1)'$  & $Z'$ & $\tan \alpha$ & $\tan \beta$ \\
          \midrule
          $U(1)_A$           & $+Z_{\cancel{d}}$ & $-2\sqrt{6}$      & $\sqrt{\tfrac{3}{5}}$  \\
          $U(1)_{21\bar{I}}$ & $+Z_{\cancel{n}}$ & $4\sqrt{\tfrac{2}{3}}$     & $-\tfrac{1}{\sqrt{7}}$ \\
          $U(1)_{21\bar{A}}$ & $+Z_{\cancel{p}}$ & $\tfrac{2\sqrt{2/3}}{3}$   & $\tfrac{1}{\sqrt{7}}$ \\
          $U(1)_{31R}$       & $-Z_{B-L}$        & $-\sqrt{\tfrac{2}{3}}$     & $0$ \\
          $U(1)_{31I}$       & $-Z_{\cancel{\ell}}$ & $\tfrac{2\sqrt{2/3}}{3}$ & $-\tfrac{3}{\sqrt{7}}$ \\
          $U(1)_{42R}$       & $-Z_{\psi}$       & $0$               & $\infty$ \\
          $U(1)_{\chi RI}$   & $+Z_{\chi}$       & $0$               & $0$ \\
          \bottomrule \bottomrule
        \end{tabular}
      \end{table}

\FloatBarrier

 \subsubsection{Leptophilic models and   \texorpdfstring{$\gamma$--$Z'$}{A-Z}  kinetic mixing}

 \label{sec:leptophilic-mixing}

       Leptophilic extensions of the SM   assign nonzero  $U(1)'$ charges   to leptons while leaving quarks  neutral at tree level.   We write the interactions for the anomaly-free symmetries $U(1)_{L_a-L_b}$, with $a,b\in\{e,\mu,\tau\}$ and $a\ne b$, to describe direct electronic scattering and the loop-induced nuclear contribution using a common charge convention~\cite{He:1991qd,Zhang:2020fiu}. The indices $a$ and $b$ select the two lepton families that define the symmetry and, correspondingly, the charged leptons $\ell_a$ and $\ell_b$ that run in the kinetic-mixing loop. We reserve $f\in\{e,\mu,\tau\}$ for the flavor of the external antineutrino. For the reactor source considered here, $f=e$. The family charges are
      \begin{equation}
        Q_f^{ab}=\delta_{fa}-\delta_{fb},
        \qquad f=e,\mu,\tau.
        \label{eq:leptophilic-family-charges}
      \end{equation}
      Thus, family $a$ has charge $+1$, family $b$ has charge $-1$, and the remaining family is neutral. The three assignments are shown in \autoref{tab:leptophilic-charges} .

      \begin{table}[ht]
        \centering

         \caption{\label{tab:leptophilic-charges}  Family  charges  under the   three $U(1)_{L_a-L_b}$ symmetries, defined in Eq. ~ \eqref{eq:leptophilic-family-charges}~ \cite{He:1991qd,Demirci:2026nju}.   Each  charged lepton  and its associated  left-handed  neutrino   carry the same family charge .}
        \begin{tabular}{lccc}
          \toprule \toprule
          Model &   $Q_e^{ab}$  &   $Q_\mu^{ab}$  &   $Q_\tau^{ab}$  \\
          \midrule
          $L_e - L_\mu$     & $+1$ & $-1$ & $0$  \\
          $L_e - L_\tau$    & $+1$ & $0$  & $-1$ \\
          $L_\mu - L_\tau$  & $0$  & $+1$ & $-1$ \\
          \bottomrule \bottomrule
        \end{tabular}

       \end{table}

      In terms of the charged-lepton fields $\ell_h$ and $P_L=(1-\gamma^5)/2$, the  interaction Lagrangian is \cite{He:1991qd,Zhang:2020fiu,Demirci:2026nju}
        \begin{equation}
        \resizebox{0.42\textwidth}{!}{$
          \mathcal L_{L_a-L_b}^{\mathrm{int}}
          =-g_{Z'}Z'_\mu\sum_{h=e,\mu,\tau}Q_h^{ab}\bigl[
            \bar\ell_h\gamma^\mu\ell_h+
            \bar\nu_h\gamma^\mu P_L\nu_h
          \bigr].
        $}
        \label{eq:leptophilic-lagrangian}
      \end{equation} 
       The overall minus sign is the same as in Eq.~\eqref{eq:zprime-interaction}, and the family charges have the signs listed in \autoref{tab:leptophilic-charges}. Comparing the two Lagrangians gives
      \begin{equation}
        \begin{aligned}
          g_V^{Z'}\boldsymbol{(}\bar\nu_f\boldsymbol{)}
          &=g_A^{Z'}\boldsymbol{(}\bar\nu_f\boldsymbol{)}=Q_f^{ab},\\
          g_V^{Z'}\boldsymbol{(}e\boldsymbol{)}&=2Q_e^{ab},
          \qquad g_A^{Z'}\boldsymbol{(}e\boldsymbol{)}=0,\\
          g_{V,A}^{Z'}\boldsymbol{(}u\boldsymbol{)}
          &=g_{V,A}^{Z'}\boldsymbol{(}d\boldsymbol{)}=0.
        \end{aligned}
        \label{eq:leptophilic-couplings}
      \end{equation}
      The factor of two in the electron vector coupling compensates for the prefactor $g_{Z'}/2$ in Eq.~\eqref{eq:zprime-interaction}. For the reactor channel $f=e$, the couplings in Eq.~\eqref{eq:leptophilic-couplings} can be inserted directly into Eqs.~\eqref{eq:eves-effective-couplings} and \eqref{eq:cross-section-zprime-eves}. These couplings yield a tree-level $Z'$ contribution to E$\nu e$S  whenever $Q_e^{ab}\ne0$.

\begin{figure}[htbp]
        \centering

        \IfFileExists{Pictures/LaLb/CEvNS_Diagram_Mix.pdf}{%
          \includegraphics[width=\linewidth]{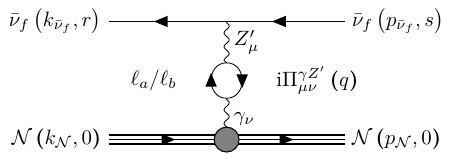}%
        }{%
          \fbox{\parbox{0.9\linewidth}{\centering Figura pendiente:\\
          \texttt{Pictures/LaLb/CEvNS\_Diagram\_Mix.pdf}}}%
        }
         \caption{\label{fig:mixed-propagator-diagram} CE$\nu \mathcal{N}$S mediated by the mixed   $\gamma$ --$Z'$ propagator. The   charged-lepton loops $\ell_a$  and  $\ell_b$, selected by  the   $L_a-L_b$ symmetry,         
         carry opposite family charges and generate the kinetic-mixing insertion through their loop contributions. The external antineutrino has flavor $f$, with $f=e$ for  the   reactor source.}

       \end{figure}

      Since quarks   are neutral under $U(1)_{L_a-L_b}$, the nuclear interaction instead arises through $\gamma$--$Z'$ kinetic mixing generated by  charged-lepton loops  as illustrated in   \autoref{fig:mixed-propagator-diagram}. We take $e=\sqrt{4\pi\alpha_{\mathrm{EM}}}\simeq0.302822>0$ and fix the electromagnetic convention by $\mathcal L_{\mathrm{EM}}=-eA_\mu J_{\mathrm{EM}}^\mu$, with charged-lepton electric charge $Q_f^{\mathrm{EM}}=-1$ and nuclear charge $Z$. The photon vertex is then $-\mathrm{i}eQ^{\mathrm{EM}}\gamma^\mu$ for a fermion. Assuming no
      independent bare kinetic mixing, we define~\cite{Zhang:2020fiu,AtzoriCorona:2022moj}
          \begin{equation}\label{eq:mixing-parameter}
            \begin{aligned}
              \epsilon_{\gamma Z'}^{ab}(q^2)
              &= \frac{e g_{Z'}}{2\pi^2} \int_0^1 dx\, x(1-x) \\
              &\quad \times \ln\left( \frac{m_b^2 - x(1-x)q^2}{m_a^2 - x(1-x)q^2} \right).
            \end{aligned}
          \end{equation}%
        
       Here $m_a$ and $m_b$ are the charged-lepton masses. This function already includes the two gauge couplings from the loop. Its full momentum dependence is retained, including when one of the charged leptons in the loop is the electron.

      The   corresponding transverse insertion defines the sign convention for the mixing:
      \begin{equation}
        \mathrm{i}\Pi_{\mu\nu}^{\gamma Z'}\boldsymbol{(}q\boldsymbol{)}
        =
        +\mathrm{i}\epsilon_{\gamma Z'}^{ab}\boldsymbol{(}q^2\boldsymbol{)}
        (q^2g_{\mu\nu}-q_\mu q_\nu).
        \label{eq:leptophilic-mixing-insertion}
      \end{equation} 
       Equivalently, it is the quadratic momentum-space kernel associated with $+\epsilon_{\gamma Z'}^{ab}Z'_{\mu\nu}F^{\mu\nu}/2$, where $Z'_{\mu\nu}$ and $F_{\mu\nu}$ are the two field strengths. Between conserved currents, the longitudinal term drops out. Attaching the free propagators to Eq.~\eqref{eq:leptophilic-mixing-insertion} then gives
      \begin{equation}
        \begin{aligned}
          D_{\mu\nu}^{Z'\gamma}
          &=
          \frac{-\mathrm{i}g_{\mu\alpha}}{q^2-m_{Z'}^2}
          \bigl[+\mathrm{i}\epsilon_{\gamma Z'}^{ab}
          \boldsymbol{(}q^2\boldsymbol{)}q^2g^{\alpha\beta}\bigr]
          \frac{-\mathrm{i}g_{\beta\nu}}{q^2}\\
          &=-\frac{\mathrm{i}\epsilon_{\gamma Z'}^{ab}
          \boldsymbol{(}q^2\boldsymbol{)}g_{\mu\nu}}{q^2-m_{Z'}^2}.
        \end{aligned}
        \label{eq:mixed-propagator}
      \end{equation}
      The denominators carry the usual $+\mathrm{i}0$ prescription, omitted here since $q^2<0$. The nuclear electromagnetic current is $j_{\mathrm{EM}}^\nu=ZF_{\mathrm{ch}}\boldsymbol{(}q^2\boldsymbol{)}(p_{\mathcal N}+k_{\mathcal N})^\nu$,  where   $F_{\mathrm{ch}}$ is the nuclear charge form factor. In the electromagnetic convention above, its vertex is $-\mathrm{i}e\,j_{\mathrm{EM}}^\nu$. With $j_{\bar\nu_f}^{\mu}=\bar v^{\,r}\boldsymbol{(}k_{\bar\nu_f}\boldsymbol{)}\gamma^\mu(1-\gamma^5)v^s\boldsymbol{(}p_{\bar\nu_f}\boldsymbol{)}$, where $k$ and $p$ denote incoming and outgoing four-momenta  and   $r,s$ are spin indices, the product of the two vertices and the mixed propagator is
      \begin{equation}
        \begin{aligned}
          \mathrm{i}\mathcal M_{L_a-L_b}^{f}
          &=
          \left(-\frac{\mathrm{i}g_{Z'}Q_f^{ab}}{2}\right)
          j_{\bar\nu_f}^{\mu}
          D_{\mu\nu}^{Z'\gamma}
          \left(-\mathrm{i}e\,j_{\mathrm{EM}}^\nu\right),\\
          \mathcal M_{L_a-L_b}^{f}
          &=
          \frac{e g_{Z'}Q_f^{ab}Z\,
          \epsilon_{\gamma Z'}^{ab}\boldsymbol{(}q^2\boldsymbol{)}}{2(q^2-m_{Z'}^2)}
          F_{\mathrm{ch}}\boldsymbol{(}q^2\boldsymbol{)}\\
          &\quad\times j_{\bar\nu_f}^{\mu}
          (p_{\mathcal N}+k_{\mathcal N})_\mu.
        \end{aligned}
        \label{eq:leptophilic-nuclear-amplitude}
      \end{equation} 
        In particular, the product $(-\mathrm{i})(-\mathrm{i})(-\mathrm{i})=\mathrm{i}$ fixes the sign before dividing by $\mathrm{i}$ to obtain $\mathcal M$. In the same external-state convention,  the   coherent SM amplitude is
      \begin{equation}
        \mathcal M_{\mathrm{SM}}^{f}
        =-\frac{G_F}{\sqrt2}Q_V^{\mathrm{SM}}
        F_W\boldsymbol{(}q^2\boldsymbol{)}
        j_{\bar\nu_f}^{\mu}(p_{\mathcal N}+k_{\mathcal N})_\mu.
        \label{eq:leptophilic-sm-amplitude}
      \end{equation}
      Here $F_W\boldsymbol{(}q^2\boldsymbol{)}$ denotes the weak form factor evaluated at $|\mathbf q|^2\simeq-q^2$. In the low-transfer regime we use $F_{\mathrm{ch}}\simeq F_W\simeq1$. Dividing  Eq.~  \eqref{eq:leptophilic-nuclear-amplitude} by Eq.~\eqref{eq:leptophilic-sm-amplitude}, and using $q^2=-2m_{\mathcal N}T_{\mathcal N}$, yields
      \begin{equation}
        \begin{aligned}
          \mathcal M_{\mathrm{tot}}^{f}
          &=\mathcal G_{L_a-L_b}^{f}
          \boldsymbol{(}T_{\mathcal N}\boldsymbol{)}
          \mathcal M_{\mathrm{SM}}^{f},\\
          \mathcal G_{L_a-L_b}^{f}
          \boldsymbol{(}T_{\mathcal N}\boldsymbol{)}
          &=1+\frac{e g_{Z'}Q_f^{ab}Z}{\sqrt2G_FQ_V^{\mathrm{SM}}}\\
          &\quad\times
          \frac{\epsilon_{\gamma Z'}^{ab}
          \boldsymbol{(}-2m_{\mathcal N}T_{\mathcal N}\boldsymbol{)}}{m_{Z'}^2+2m_{\mathcal N}T_{\mathcal N}}.
        \end{aligned}
        \label{eq:leptophilic-cevns-rescaling}
      \end{equation} 
        This normalization uses the signed weak charge of Eq.~\eqref{eq:sm-nuclear-charge}. In particular, the coefficient is $1/(\sqrt2G_F)$, with the source-flavor charge $Q_f^{ab}$ retained explicitly. The shift $Q_V^{\mathrm{SM}}(\mathcal G_{L_a-L_b}^{f}-1)$ agrees with the loop-induced proton contribution in Ref.~\cite{AtzoriCorona:2022moj}. The differential cross section is therefore
      \begin{equation}
        \left.\frac{d\sigma_{\bar\nu_f\mathcal N}}{dT_{\mathcal N}}\right|_{\mathrm{SM}+L_a-L_b}
        =
        \left|\mathcal G_{L_a-L_b}^{f}
        \boldsymbol{(}T_{\mathcal N}\boldsymbol{)}\right|^2
        \left.\frac{d\sigma_{\bar\nu_f\mathcal N}}{dT_{\mathcal N}}\right|_{\mathrm{SM}}.
        \label{eq:cross-section-leptophilic-cevns}
      \end{equation} 
        Under a simultaneous reversal of all family charges, both $Q_f^{ab}$  and   $\epsilon_{\gamma Z'}^{ab}$ change sign, leaving their product and the interference unchanged. The SM limit follows from $g_{Z'}\to0$, since the amplitude correction is proportional to $e^2g_{Z'}^2$, and the nuclear mediator contribution vanishes for $Q_f^{ab}=0$. The sign of the interference is determined relative to the signed $Q_V^{\mathrm{SM}}$, not by the explicit plus sign in Eq.~\eqref{eq:leptophilic-cevns-rescaling} alone.

        For a reactor source, $f=e$. The charges in \autoref{tab:leptophilic-charges} imply identical direct E$\nu e$S cross sections for $L_e-L_\mu$ and $L_e-L_\tau$ at fixed $g_{Z'}$ and $m_{Z'}$, whereas their nuclear contributions differ through the masses entering Eq.~\eqref{eq:mixing-parameter}. For $L_\mu-L_\tau$, $Q_f^{\mu\tau}=Q_e^{\mu\tau}=0$, so neither of these additional diagrams contributes for the incident $\bar\nu_f=\bar\nu_e$. This statement concerns the perturbative contributions retained here: tree-level E$\nu e$S and  CE$\nu \mathcal{N}$S   through one mixing insertion. Loop corrections to the direct electron vertex and an independent bare kinetic mixing are not included .

\FloatBarrier
\section{Statistical Analysis}
\label{sec:analysis}
  In this section, we present the statistical analysis of \textsc{Conus}$+$ data used to constrain the scenarios under study in this work. We closely follow the methodology applied in Ref.~\cite{DeRomeri:2025csu}. We also present in detail the main experimental characteristics and the required parameters.

\subsection{The \texorpdfstring{\textsc{Conus}$+$}{CONUS+} Experiment}
\label{sec:experiment}

     The \textsc{Conus}$+$ experiment has reported the first observation of CE$\nu \mathcal{N}$S with antineutrinos, with a statistical significance of 3.7$\sigma$~\cite{Ackermann:2025obx}. For detection, \textsc{Conus}$+$ used high-purity germanium detectors with kilogram-scale masses and very low keV-range thresholds. These detectors are located 20.7 m from the Leibstadt nuclear reactor in Switzerland. This discovery enables precision tests of standard-model parameters, such as the weak mixing angle, and the exploration of new physical scenarios beyond the standard model, as considered in this work.

    The first reactor-antineutrino CE$\nu \mathcal{N}$S analysis combines the spectra of three HPGe detectors, denoted C2, C3, and C5. The fourth detector, C4, was not included in the final dataset because of rate instabilities reported by the collaboration. \cite{Ackermann:2025obx} Each detector has its own ionization threshold, active mass, and reactor-ON exposure; these detector-dependent inputs determine the signal normalization and the accessible reconstructed-energy range. The values used in the present calculation are collected in   \autoref{tab:conus-parameters} .

    \begin{table}[ht]
      \centering
      \caption{\label{tab:conus-parameters}Experimental parameters of the \textsc{Conus}$+$ experiment used in the statistical
      analysis. The data correspond to the first direct observation of coherent
      elastic antineutrino--nucleus scattering (CE$\nu \mathcal{N}$S)~\cite{Ackermann:2025obx}.}
      \begin{tabular*}{\columnwidth}{@{\extracolsep{\fill}}ll}
        \toprule \toprule
        \textbf{Parameter} & \textbf{Value} \\
        \midrule

        \multicolumn{2}{l}{\textit{Detector system}} \\
        \quad Thresholds & 180/160/170~eV$_{\mathrm{ee}}$ \\
        \quad Active mass & 0.95/0.94/0.94~kg\\
        \quad Exposure & 117/110/119~days \\

        \addlinespace[0.5ex]

        \multicolumn{2}{l}{\textit{Antineutrino source}} \\
        \quad Antineutrino flux & {\scriptsize  $1.5 \times 10^{13}$ cm$^{-2}$ s$^{-1}$} \\
        \quad Thermal power $P_{\mathrm{th}}$ & $3.6$~GW \\
        \quad Fission fractions & 53-8-32-7~\% \\

        \addlinespace[0.5ex]

        \multicolumn{2}{l}{\textit{Observational result}} \\
        \quad Events (measured) & $(395 \pm 106)$ \\
        \quad Events (SM) & $(347 \pm 59)$ \\
        \quad Data/prediction & $1.14 \pm 0.36$ \\
        \quad SM prediction uncertainty & $17\%$ \\

        \bottomrule \bottomrule
      \end{tabular*}
    \end{table}

    These inputs   determine  the   nuclear and electronic signal contributions discussed below, expressed in the same normalization as the measured excess spectrum .

\subsection{  Nuclear and electronic  signal   prediction}
\label{sec:signal-prediction}
    The signal prediction combines nuclear and electronic recoils in the same reconstructed-energy bins. The published excess is obtained by comparing reactor-on data with a background model constrained by reactor-off measurements~\cite{Ackermann:2025obx}. We therefore treat reactor-induced E$\nu e$S as part of the signal prediction, rather than as an independently subtracted background. Both contributions use the same exposure and flux normalization and are expressed in the normalization of the measured excess spectrum. The nuclear contribution in bin  $i$  is given by
      \begin{equation}
      R_{i}^{\mathrm{CE\nu\mathcal N S}} =
      \int_{E_{i}^{\text{low}}}^{E_{i}^{\text{high}}}
      \frac{dR_{\mathrm{CE\nu\mathcal N S}}}{dE_{\text{ee}}^{\text{reco}}}
      \, dE_{\text{ee}}^{\text{reco}},
      \label{eq:nuclear-bin-prediction}
    \end{equation}

     where the bin edges $E_{i}^{\text{low}}, E_{i}^{\text{high}}$ are given in electron-equivalent energy (eV$_{\text{ee}}$). The reconstructed-energy bins span the interval $160$--$350~\text{eV}_{\text{ee}}$ in steps of $10~\text{eV}_{\text{ee}}$. \cite{DeRomeri:2025csu} The observable quantity is the reconstructed electron-equivalent energy $E_{\text{ee}}^{\text{reco}}$, which is related to the underlying nuclear recoil energy $T_{\mathcal{N}}$ through detector effects. Since the CE$\nu \mathcal{N}$S cross section is fundamentally expressed in terms of $T_{\mathcal{N}}$, the event rate is computed in this variable and mapped to the observable through the detector response function. Accordingly, the differential event rate is written as

    \begin{equation}
        \begin{aligned}
          \frac{dR_{\mathrm{CE\nu\mathcal N S}}}{dE_{\text{ee}}^{\text{reco}}} =\; & \mathcal{E}
          \int_{T_{\mathcal{N}}^{\text{min}}}^{T_{\mathcal{N}}^{\text{max}}}
          dT_{\mathcal{N}} \;
          \mathcal{F}(T_{\mathcal{N}}) \, \mathcal{G}(E_{\text{ee}}^{\text{reco}}, T_{\mathcal{N}}) \\
          & \times
          \int_{E_{\bar\nu_e}^{\text{min}}(T_{\mathcal{N}})}^{E_{\bar\nu_e}^{\text{max}}}
          dE_{\bar\nu_e}\;
          \frac{d\Phi_{\bar\nu_e}}{dE_{\bar\nu_e}}\;
          \frac{d\sigma_{\bar\nu_e\mathcal{N}}}{dT_{\mathcal{N}}},
        \end{aligned}
      \label{eq:nuclear-differential-rate}
    \end{equation}

     where $\mathcal{E}$ denotes the detector exposure, defined as the product of the effective data-taking time and the number of target nuclei contained in the active detector volume. In the present implementation, the detection efficiency is taken to be effectively unity within the integration domain, in accordance with the \textsc{Conus}$+$ specifications~\cite{Ackermann:2025obx}, which report an efficiency $\mathcal{F}(T_{\mathcal{N}})$ close to $100\%$ above the experimental threshold.

    The detector-dependent thresholds, active masses, and exposures listed in   \autoref{tab:conus-parameters}  enter the computation of $R_i^{\text{th}}$ through the exposure factor and the lower accessible reconstructed-energy range. The same table also summarizes the total \textsc{Conus}$+$ event excess, $(395\pm106)$, and the corresponding SM expectation, $(347\pm59)$, which provide the global normalization reference for the spectral fits discussed below.

    The detector response function $\mathcal{G}$ accounts for the stochastic mapping between the nuclear recoil energy and the reconstructed ionization signal and is modeled as a Gaussian distribution centered at the mean ionization energy~\cite{Lindner:2024eng,Chattaraj:2025fvx}.

    \begin{equation}
      \mathcal{G}(E_{\text{ee}}^{\text{reco}}, T_{\mathcal{N}}) =
      \frac{\exp\!\left[
      -\frac{\left(E_{\text{ee}}^{\text{reco}} - E_{\text{er}}(T_{\mathcal{N}})\right)^2}
      {2\sigma_{\text{res}}^2(T_{\mathcal{N}})}
      \right]}{\sqrt{2\pi}\,\sigma_{\text{res}}(T_{\mathcal{N}})}
      .
      \label{eq:nuclear-response}
    \end{equation}

    The energy resolution is parameterized as

    \begin{equation}
      \sigma_{\text{res}}(T_{\mathcal{N}}) =
      \sqrt{\sigma_0^2 + F_{\text{Fano}}\,\eta\,E_{\text{er}}(T_{\mathcal{N}})},
      \label{eq:energy-resolution}
    \end{equation}

     where $\sigma_0 = 20.38~\text{eV}_{\text{ee}}$, $F_{\text{Fano}} = 0.1096$, and $\eta = 2.96~\text{eV}_{\text{ee}}$ for germanium detectors~\cite{Ackermann:2025obx,CONUS:2024lnu}.

    The mean ionization energy $E_{\text{er}}$ is related to the nuclear recoil energy $T_{\mathcal{N}}$ through the Lindhard quenching factor,

    \begin{equation}
      E_{\text{er}}(T_{\mathcal{N}}) =
      Q_F(T_{\mathcal{N}})\, T_{\mathcal{N}},
      \label{eq:ionization-energy}
    \end{equation} 
    where   $E_{\mathrm{er}}$ and  $T_{\mathcal{N}}$   are expressed in the same energy  units. The quenching factor is given by~\cite{Rink:2022rsx, Bonhomme:2022lcz}
      \begin{equation}
      Q_F(T_{\mathcal{N}}) =
      \frac{k\,g(\epsilon)}{1 + k\,g(\epsilon)},
      \qquad
      k = 0.162,
      \label{eq:lindhard-quenching}
    \end{equation} 
    with $ g(\epsilon) = 3\,\epsilon^{0.15} + 0.7\,\epsilon^{0.6} + \epsilon$ and   $\epsilon = 11.5\,Z^{-7/3}\,(T_{\mathcal{N}}/\mathrm{keV})$ .

    The lower reconstructed-energy threshold is detector dependent and is given by the ionization thresholds listed in   \autoref{tab:conus-parameters} . In the combined spectral analysis, the reconstructed bins start at $160~\text{eV}_{\text{ee}}$, corresponding to the lowest threshold among the detectors included in the dataset. Accordingly, the lower reconstructed-energy acceptance is imposed by the bin edges and detector thresholds, while the recoil-energy integration limits are fixed by the numerical integration domain and by kinematics. In the present implementation, the lower recoil cutoff is taken as $T_{\mathcal{N}}^{\text{min}} = 2.72 \times 10^{-5}~\text{MeV}$, well below the reconstructed-energy analysis window.

    The maximum recoil energy is fixed by kinematics as

    \begin{equation}
      T_{\mathcal{N}}^{\text{max}} =
      \frac{2 (E_{\bar\nu_e}^{\text{max}})^{2}}{m_{\mathcal{N}} + 2 E_{\bar\nu_e}^{\text{max}}},
      \label{eq:maximum-nuclear-recoil}
    \end{equation} 
    the minimum   antineutrino  energy required to produce a recoil $T_{\mathcal{N}}$ is given by
      \begin{equation}
      E_{\bar\nu_e}^{\text{min}}(T_{\mathcal{N}}) =
      \frac{T_{\mathcal{N}}}{2} +
      \sqrt{\left(\frac{T_{\mathcal{N}}}{2}\right)^2 +
      \frac{m_{\mathcal{N}} T_{\mathcal{N}}}{2}},
      \label{eq:minimum-antineutrino-energy}
    \end{equation} 
    while the upper limit is fixed to   $    E_{\bar\nu_e}^{\text{max}} = 10~\text{MeV}$ .

    The reactor antineutrino flux is modeled as a superposition of the contributions from the dominant fissile isotopes $l = \{^{235}\mathrm{U},\,^{238}\mathrm{U},\,^{239}\mathrm{Pu},\,^{241}\mathrm{Pu}\}$~\cite{Ackermann:2025obx}. The differential flux at the detector is written as

    \begin{equation}
    \resizebox{0.42\textwidth}{!}{$
      \frac{d\Phi_{\bar\nu_e}}{d E_{\bar\nu_e}}(E_{\bar\nu_e},t)=
      \frac{P_{\mathrm{th}}(t)}{4\pi L^2\sum_l f_l(t)E_l^{\mathrm{fis}}}
      \sum_l f_l(t)\,\lambda_l(E_{\bar\nu_e}).
    $}
    \label{eq:differential-flux-expanded}
    \end{equation}

    Here $P_{\mathrm{th}}$ is the reactor thermal power, $L$ is the reactor--detector baseline, $f_l$ are the fission fractions, $E_l^{\mathrm{fis}}$ are the energies released per fission, and   $\lambda_l(E_{\bar\nu_e})$  is the antineutrino spectrum per fission reported in Ref.~\cite{Baldoncini:2014vda, Zhang:2023zif}. The numerical inputs adopted in this work are listed in   \autoref{tab:conus-parameters} . For the time-integrated \textsc{Conus}$+$ dataset, $P_{\mathrm{th}}$ and $f_l$ are treated as effective constants.

    For   antineutrino  energies above 2 MeV, we followed Ref.~\cite{Mueller:2011nm}, where each isotopic spectrum is described by a phenomenological parameterization,
      \begin{equation}
    \lambda_l(E_{\bar\nu_e}) = \exp\!\left[\sum_{k=1}^{6} \alpha_{lk}\,E_{\bar\nu_e}^{k-1}\right].
      \label{eq:isotope-spectrum}
    \end{equation} 

    The coefficients $\alpha_{lk}$ are obtained from a fit to the converted $\beta$-spectra using the MIGRAD minimization algorithm within the \texttt{TMinuit} framework.\footnote{The   antineutrino  energy   $E_{\bar\nu_e}$  is expressed in MeV in the above parametrization.} The central values of the coefficients $\alpha_{lk}$ reported in Ref.~\cite{Mueller:2011nm, Baldoncini:2014vda} are adopted, and their correlations are neglected. For   antineutrino  energies below 2 MeV, we   use the spectrum of  Ref.~\cite{Kopeikin:2012zz} .

      The electronic contribution uses the free-electron cross sections in \eqref{eq:cross-section-sm-eves} and \eqref{eq:cross-section-zprime-eves}, weighted by the effective number of ionizable electrons per germanium atom. With $T_e$ the true electronic recoil energy, it is
    \begin{equation}
      \begin{aligned}
        R_i^{\mathrm{E\nu eS}}={}&\mathcal E
        \int_{E_i^{\mathrm{low}}}^{E_i^{\mathrm{high}}}dE_{\mathrm{ee}}^{\mathrm{reco}}
        \int_0^{T_e^{\max}}dT_e\,
        Z_{\mathrm{eff}}^{\mathrm{Ge}}\boldsymbol{(}T_e\boldsymbol{)}\\
        &\times\mathcal G_e\boldsymbol{(}E_{\mathrm{ee}}^{\mathrm{reco}},T_e\boldsymbol{)}\\
        &\times\int_{E_{\bar\nu_e}^{\min}(T_e)}^{E_{\bar\nu_e}^{\max}}
        dE_{\bar\nu_e}\,
        \frac{d\Phi_{\bar\nu_e}}{dE_{\bar\nu_e}}\,
        \frac{d\sigma_{\bar\nu_e e}}{dT_e}.
      \end{aligned}
      \label{eq:electron-bin-prediction}
    \end{equation} 
      Here $\mathcal E$ counts target atoms times exposure, as in Eq.~\eqref{eq:nuclear-differential-rate}, so no additional factor of $Z$ multiplies $Z_{\mathrm{eff}}^{\mathrm{Ge}}$. We use the ten-step germanium prescription in Table~II of Ref.~\cite{AtzoriCorona:2022dresden}; at a binding-energy boundary, the lower electron count is retained. The Gaussian response $\mathcal G_e$ has mean $E_{\mathrm{er}}=T_e$ (unit electronic quenching) and the same resolution parameters as the nuclear response. The kinematic limits follow from Eqs.~\eqref{eq:maximum-nuclear-recoil} and \eqref{eq:minimum-antineutrino-energy} with $m_{\mathcal N}\to m_e$  and   $T_{\mathcal N}\to T_e$, with $E_{\bar\nu_e}^{\max}=10$ MeV. The resulting total prediction is
    \begin{equation}
      R_i^{\mathrm{th}}=R_i^{\mathrm{CE\nu\mathcal N S}}+R_i^{\mathrm{E\nu eS}}.
      \label{eq:total-bin-prediction}
    \end{equation}
    In \eqref{eq:total-bin-prediction}, each channel includes its SM contribution and, where applicable, the mediator contribution and its interference with the SM. For the quark NSI considered here, the electronic contribution remains the SM one. The distinct nuclear and electronic final states are summed as rates, with the SM term included once in each channel; no interference between the two channels is added. Thus, \eqref{eq:electron-bin-prediction} introduces neither double counting nor additional electron-NSI parameters.

  \subsection{Profiled statistic and numerical comparisons}
\label{sec:profiled-analysis}
    We constrain the model parameters $\boldsymbol\theta$ by fitting the total prediction in \eqref{eq:total-bin-prediction} to the 19-bin excess spectrum. The normalization nuisance parameter $\alpha$ multiplies both signal components together. We use
    \begin{equation}
      \chi^2\boldsymbol{(}\boldsymbol\theta,\alpha\boldsymbol{)}=
      \sum_{i=1}^{19}\frac{[R_i^{\mathrm{exp}}-(1+\alpha)R_i^{\mathrm{th}}]^2}{\sigma_i^2}
      +\frac{\alpha^2}{\sigma_\alpha^2},
      \label{eq:profiled-chi-square}
    \end{equation}
    where $\sigma_i$ is the average of the reported upper and lower bin uncertainties. The Gaussian normalization uncertainty is $\sigma_\alpha=0.169$, incorporating  flux (4.6\%), quenching (7.3\%), threshold (14.1\%), active mass (1.1\%), trigger efficiency (0.7\%), and   weak-form-factor  (3.2\%)  contributions ~\cite{Ackermann:2025obx}.  This is an effective common normalization prescription: channel-dependent effects, such as nuclear quenching, are not assigned separate nuisance parameters or propagated as spectral distortions. The same prescription is used in the nuclear-only and combined fits. At each model point, the unrestricted Gaussian profile is obtained analytically:
    \begin{equation}
      \hat\alpha\boldsymbol{(}\boldsymbol\theta\boldsymbol{)}=
      \frac{\sum_i R_i^{\mathrm{th}}(R_i^{\mathrm{exp}}-R_i^{\mathrm{th}})/\sigma_i^2}{\sum_i (R_i^{\mathrm{th}})^2/\sigma_i^2+\sigma_\alpha^{-2}}.
      \label{eq:analytic-normalization}
    \end{equation}
    The normalization parameter in \eqref{eq:analytic-normalization} is distinct from the angular coordinate $\alpha$ of the $E_6$ parameterization.

     For each scenario, we minimize the profiled statistic independently and define
    \begin{equation}
      \Delta\chi^2\boldsymbol{(}\boldsymbol\theta\boldsymbol{)}=
      \chi^2\boldsymbol{(}\boldsymbol\theta,\hat\alpha\boldsymbol{)}-\chi^2_{\min}.
      \label{eq:delta-chi-square-analysis}
    \end{equation}
    We adopt the nominal Wilks thresholds $1,4,9$ for the one-parameter $1\sigma$, $2\sigma$, and $3\sigma$ intervals, and $4.60517$ for two-parameter regions at 90\% C.L.~\cite{Wilks:1938dza} The latter threshold is also used for the mass--coupling boundaries. These regions use the stated asymptotic prescription; their coverage is not separately calibrated by simulation.

    To quantify the effect of E$\nu e$S, we repeat the fits with the nuclear signal alone, retaining the same data, domains, and normalization prior. Each prediction has its own $\hat\alpha$ and $\chi^2_{\min}$ in \eqref{eq:delta-chi-square-analysis}. Scans use 200 points per parameter axis. One-dimensional interval endpoints and mass--coupling crossings are linearly interpolated between adjacent grid values; quoted two-dimensional projections are extrema of allowed grid points. For values quoted at representative masses between sampled boundaries, we interpolate $\log g_{Z'}^{\mathrm{lim}}$ linearly in $\log m_{Z'}$. Graphical smoothing is not used to determine the numerical limits.

    For the NSI and Sanson--Flamsteed regions, we summarize their overlap through
    \begin{equation}
      J=\frac{\mathcal A(A\cap B)}{\mathcal A(A\cup B)},\qquad
      \delta_{\mathcal A}=100\left[\frac{\mathcal A(B)}{\mathcal A(A)}-1\right],
      \label{eq:region-comparison}
    \end{equation}
    where $A$ and $B$ are the nuclear-only and combined allowed sets in the same scanned domain. In \eqref{eq:region-comparison}, $\mathcal A$ denotes the grid estimate of area; SF areas use the projected-plane weights for $x=\alpha\cos\beta$, $y=\beta$. These measures describe agreement at the evaluated resolution, not exact equality of continuous contours or a statistical equivalence test. One-dimensional NSI intervals are compared branch by branch, rather than by subtracting coordinates of arbitrarily selected degenerate minima.

    To characterize a low-mass plateau, we take the median limit at the first five mass points and retain the initial segment within 5\% of that value. The intermediate nuclear-dominated shoulder of an $E_6$ combined boundary is identified where the limit is within 5\% of both the nuclear-only plateau and the nuclear-only boundary at the same mass. These tolerances describe the numerical shape of a boundary; they are not additional confidence levels.

 \FloatBarrier  
\section{Analysis and Results}
\label{sec:results}

  In this section, we present the main results of our analysis. First, we analyze the compatibility between \textsc{Conus}$+$ data and the theoretical predictions considered in this study through a bin-by-bin pull study. Then, we use the \textsc{Conus}$+$ CE$\nu \mathcal{N}$S data to constrain BSM scenarios such as neutrino nonstandard interactions,   $E_6$ -motivated $Z'$ models, and the leptophilic   $L_e-L_\mu$ and $L_e-L_\tau$ scenarios. The reported fit statistics, intervals, projections, and mediator limits use the combined CE$\nu \mathcal{N}$S+E$\nu e$S prediction. Nuclear-only results are used to quantify the effect of including electronic scattering.

\subsection{Analysis of Pulls}
\label{sec:pull-results}
    Before deriving parameter bounds, we first examine the residual structure of the spectral fit. This bin-by-bin comparison provides a diagnostic of whether the \textsc{Conus}$+$ data prefer a coherent distortion of the CE$\nu \mathcal{N}$S recoil spectrum or are compatible with localized fluctuations around the SM expectation. Here we compare the SM with extensions including NSI, $E_6$-motivated $Z'$ scenarios, and the leptophilic $L_e\!-\!L_\mu$ model, as well as the $L_e\!-\!L_\tau$ model, using the combined CE$\nu\mathcal N$S+E$\nu e$S prediction.
    \begin{figure*}[htbp]
      \centering
      \IfFileExists{Pictures/Pulls/Plot_Pulls_Profiled_TheoryBase.pdf}{%
        \includegraphics[width=0.7\linewidth]{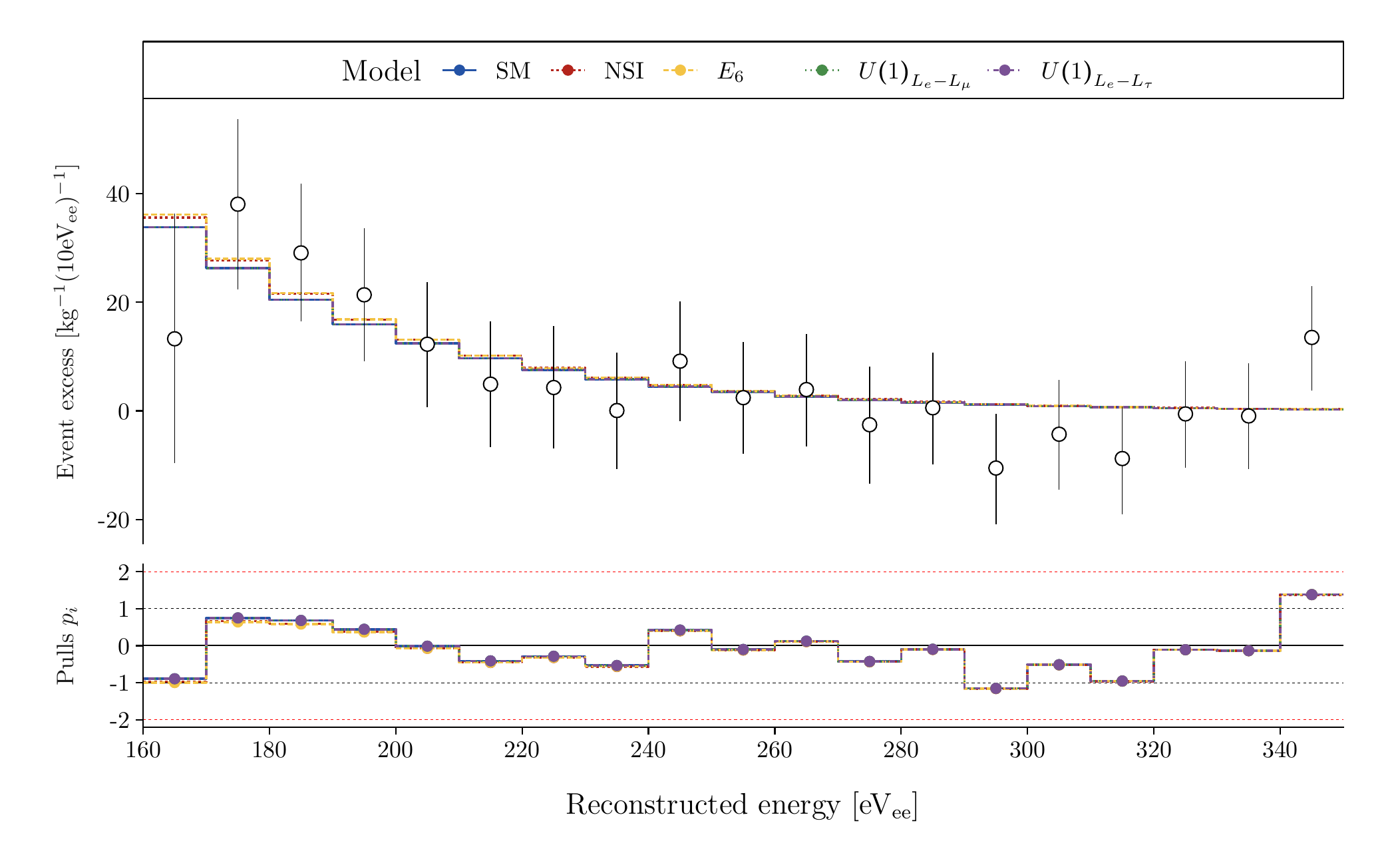}%
      }{%
        \fbox{\parbox[c][5cm][c]{0.85\linewidth}{\centering Figura pendiente:\\ \texttt{\detokenize{Plot_Pulls_Profiled_TheoryBase.pdf}}}}%
      }
      \caption{\label{fig:pulls-models}Spectral comparison and bin-by-bin pulls for the profiled models in the \textsc{Conus}$+$ excess-spectrum analysis.~\cite{Ackermann:2025obx} The upper panel shows the measured excess spectrum together with the profiled predictions for the SM, NSI, $E_6$, $L_e\!-\!L_\mu$, and $L_e\!-\!L_\tau$ hypotheses. Each prediction includes CE$\nu\mathcal N$S and E$\nu e$S, while the lower panel shows the corresponding pulls.}
    \end{figure*}
    
    The pull in each bin is defined as
    \begin{equation}
      p_i = \frac{R_i^{\mathrm{exp}}-(1+\hat\alpha)R_i^{\mathrm{th}}}{\sigma_i},
      \label{eq:bin-pull}
    \end{equation}
    where $\hat\alpha$ is the profiled normalization parameter for the corresponding model. Positive and negative values indicate that the experimental point lies above and below the profiled theoretical prediction, respectively.
    
    Fig.~\ref{fig:pulls-models} summarizes this comparison. The upper panel shows the \textsc{Conus}$+$ excess spectrum together with the profiled reconstructed predictions for the SM, NSI, $E_6$, and $L_e\!-\!L_\mu$ hypotheses, together with the $L_e\!-\!L_\tau$ prediction, while the lower panel shows the corresponding bin-by-bin pulls. With the convention above, the lower panel shows a clear but mild structure: bins $2$--$4$ are predominantly positive, bins $5$--$13$ fluctuate around zero, bins $1,14$--$18$ are mostly negative, and bin $19$ returns to a positive value. Most deviations remain at approximately the $\mathcal{O}(1\sigma)$ level, so the residuals do not indicate a coherent spectral distortion.
    
    The global statistics in Table~\ref{tab:pulls-summary} should be interpreted together with the pull morphology in Fig.~\ref{fig:pulls-models}: most bins satisfy $|p_i|\lesssim 1$, with the largest deviations concentrated in bins $14$, $16$, and $19$ at the level of $|p_i|\sim0.95$--$1.38$. At the level of absolute minimum, NSI and $E_6$ improve only marginally with respect to SM ($\chi^2_{\min}=7.2270$ and $7.1995$, respectively, compared with $7.3499$ for the SM), while both leptophilic models approach the SM value $\chi^2_{\min}=7.3499$. Thus, the largest improvement is only $\Delta\chi^2=0.1503$ for the $E_6$ fit. This marginal variation indicates that the data do not show a statistically meaningful preference for a BSM hypothesis.
    
    \begin{table}[htbp]
      \centering
      \caption{\label{tab:pulls-summary}Global fit statistics for the SM, NSI, $E_6$, and the two leptophilic models using the combined CE$\nu\mathcal N$S+E$\nu e$S prediction. The displayed degrees of freedom follow the nominal count of 19 bins minus the fitted model parameters and one normalization nuisance.}
      \begin{tabular}{lrrrr}
        \toprule \toprule
        Model & \multicolumn{1}{c}{$\sum_i p_i$} & \multicolumn{1}{c}{$\chi^2_{\min}$} & \multicolumn{1}{c}{$\chi^2_{\min}/\mathrm{dof}$} & dof \\
        \midrule
        SM & $-1.5242$ & $7.3499$ & $0.4083$ & 18 \\
        NSI & $-2.4775$ & $7.2270$ & $0.5162$ & 14 \\
        $E_6$ & $-2.4895$ & $7.1995$ & $0.5143$ & 14 \\
        $L_e\!-\!L_\mu$ & $-1.5242$ & $7.3499$ & $0.4594$ & 16 \\
        $L_e\!-\!L_\tau$ & $-1.5242$ & $7.3499$ & $0.4594$ & 16 \\
        \bottomrule \bottomrule
      \end{tabular}
    \end{table}
    
    From a goodness-of-fit perspective that also accounts for model complexity, the comparison in $\chi^2_{\min}/\mathrm{dof}$ is instructive: for the SM, $\chi^2_{\min}/\mathrm{dof}=0.4083$; for $L_e\!-\!L_\mu$, $\chi^2_{\min}/\mathrm{dof}=0.4594$; and for NSI/$E_6$, $\chi^2_{\min}/\mathrm{dof}=0.5162/0.5143$. Since the reduction in $\chi^2_{\min}$ is negligible, but the number of fitted parameters increases (smaller dof), the more complex models do not provide a commensurate gain in descriptive power. The reduced statistic is reported only as a descriptive quantity and is not used as a model-selection test. Therefore, under a parsimony criterion, the SM remains the preferred benchmark for the \textsc{Conus}$+$ pull dataset, while BSM scenarios should be interpreted as compatible alternatives rather than statistically favored explanations.
    
    From the physics standpoint, the sign structure of the pulls is also informative. Negative pulls in the high-energy tail indicate that the corresponding profiled prediction is locally above the measured excess, whereas positive pulls in the lowest and final bins indicate local underprediction. Since this alternation does not develop into a coherent, same-sign trend across the full spectrum and the magnitudes remain at the $\mathcal{O}(1\sigma)$ level, the observed residuals are better interpreted as localized shape tensions and statistical fluctuations than as evidence for a new interaction component.
    
    For completeness, adding E$\nu e$S changes $\chi^2_{\min}$ by at most $1.32\times10^{-3}$, the fitted normalization by at most $7.32\times10^{-5}$, and the largest binwise pull by only $1.05\times10^{-3}$. Therefore, the residual interpretation around the best-fit point is practically unchanged by the electronic channel. This local stability does not imply that the exclusion boundaries away from the best-fit region are insensitive to E$\nu e$S. \subsection{Neutrino Non-Standard Interactions}

\label{sec:nsi-results}
    We use quark NSI as a validation benchmark and as a reference for the mediator analysis. Their electronic signal is purely SM, so the comparison isolates the effect of adding this small component to the nuclear prediction. The combined one-parameter intervals are listed in \autoref{tab:nsi-comparison-conus-plus}, together with published comparisons.

     \begin{table*}[htbp]
      \centering

       \caption{$1\sigma$ vector NSI intervals. This work uses the combined CE$\nu\mathcal N$S+E$\nu e$S prediction, with SM electron scattering and quark NSI. The remaining columns retain the published comparison values.}
      \label{tab:nsi-comparison-conus-plus}

       \begin{tabular}{cccc}
        \toprule \toprule
        \textbf{Parameter} &
        \textbf{This work} &
        \shortstack{\textbf{\textsc{Conus}$+$}\\\cite{DeRomeri:2025csu}} &
        \shortstack{\textbf{COHERENT}\\\textbf{CsI+LAr}\\\cite{DeRomeri:2022twg}}
        \\
        \midrule
        $\varepsilon^{uV}_{ee}$
        &
          $[-0.0434,0.0207]\cup[0.3541,0.4182]$
         &
        $[-0.037,0.026]\cup[0.348,0.411]$
        &
        $[-0.024,0.045]\cup[0.34,0.43]$
        \\

        $\varepsilon^{dV}_{ee}$
        &
          $[-0.0401,0.0191]\cup[0.3269,0.3861]$
         &
        $[-0.034,0.024]\cup[0.322,0.380]$
        &
        $[-0.027,0.048]\cup[0.30,0.39]$
        \\

        $\varepsilon^{uV}_{e\mu/\tau}$
        &
          $[-0.1348,0.1348]$
         &
        $[-0.123,0.123]$
        &
        \begin{tabular}[c]{@{}c@{}}
          $e\mu:\ [-0.081,0.081]$\\
          $e\tau:\ [-0.13,0.13]$
        \end{tabular}
        \\

        $\varepsilon^{dV}_{e\mu/\tau}$
        &
          $[-0.1245,0.1245]$
         &
        $[-0.114,0.114]$
        &
        \begin{tabular}[c]{@{}c@{}}
          $e\mu:\ [-0.071,0.071]$\\
          $e\tau:\ [-0.12,0.12]$
        \end{tabular}
        \\
        \bottomrule \bottomrule
      \end{tabular}

     \end{table*}   

     \begin{figure}[htbp]
      \centering

      \IfFileExists{Pictures/NSI/Plot_x2_NSI_1D.pdf}{%
        \includegraphics[width=\linewidth]{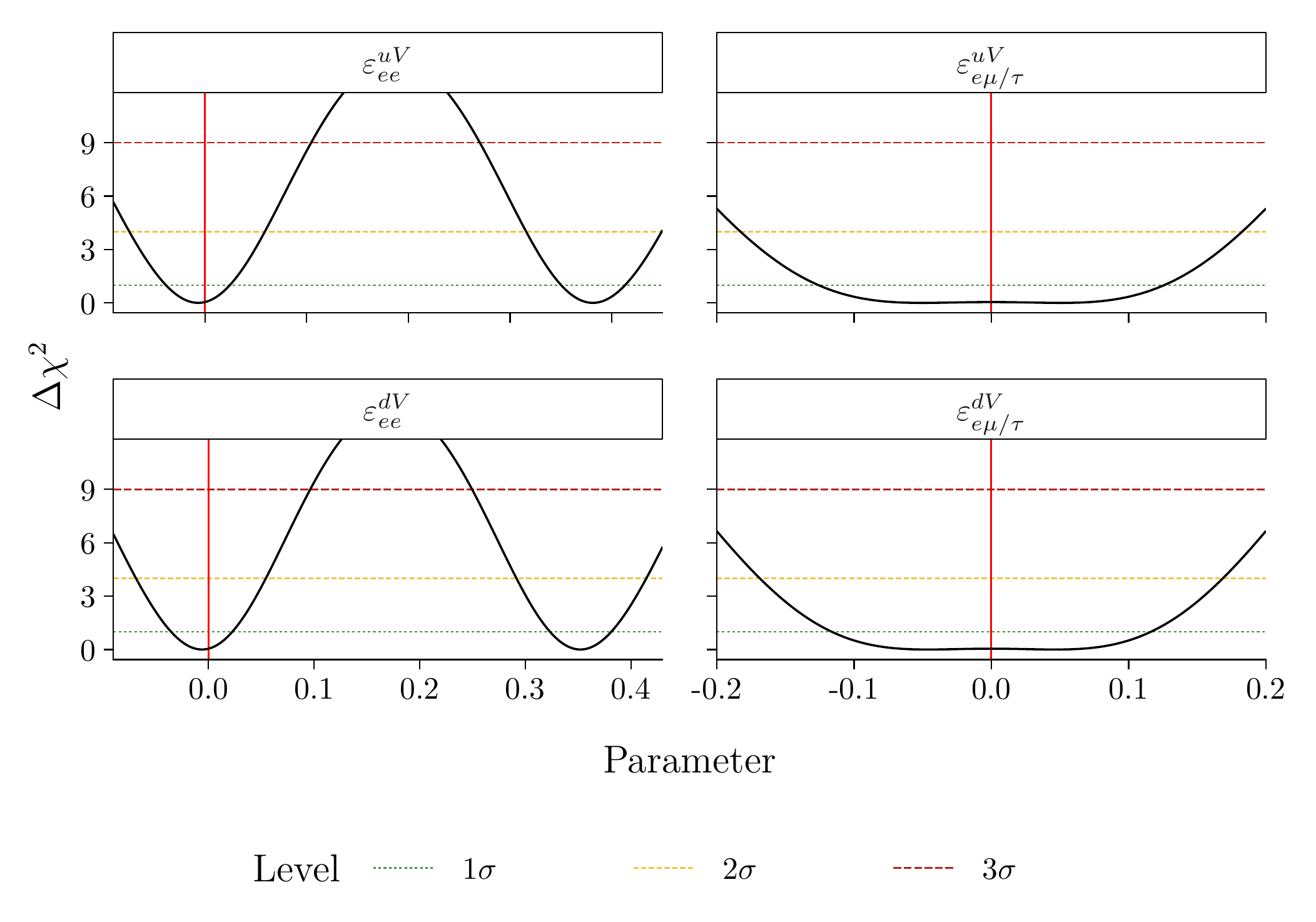}%
      }{%
        \fbox{\parbox[c][5cm][c]{0.85\linewidth}{\centering Figura pendiente:\\ \texttt{\detokenize{Plot_x2_NSI_1D.pdf}}}}%
      }
      \caption{\label{fig:nsi-one-parameter}One-dimensional $\Delta\chi^2$ profiles for individual vector NSI parameters $\varepsilon^{qV}_{\alpha\beta}$ obtained from the \textsc{Conus}$+$ CE$\nu \mathcal{N}$S + E$\nu e$S analysis.The effect of adding E$\nu e$S is quantified in the text Each parameter is varied independently while profiling over the nuisance parameter $\alpha$. The red, green, and blue dashed horizontal lines indicate the $3\sigma$, $2\sigma$, and $1\sigma$ confidence thresholds ($\Delta\chi^2=9.00,\,4.00,\,1.00$), respectively. The red vertical line marks the best-fit point for each parameter. The SM corresponds to $\varepsilon^{qV}_{\alpha\beta}=0$.}
     \end{figure}

\begin{figure}[htbp]
      \centering
      \IfFileExists{Pictures/NSI/All_Plot_x2_NSI_Region.pdf}{%
        \includegraphics[width=\linewidth]{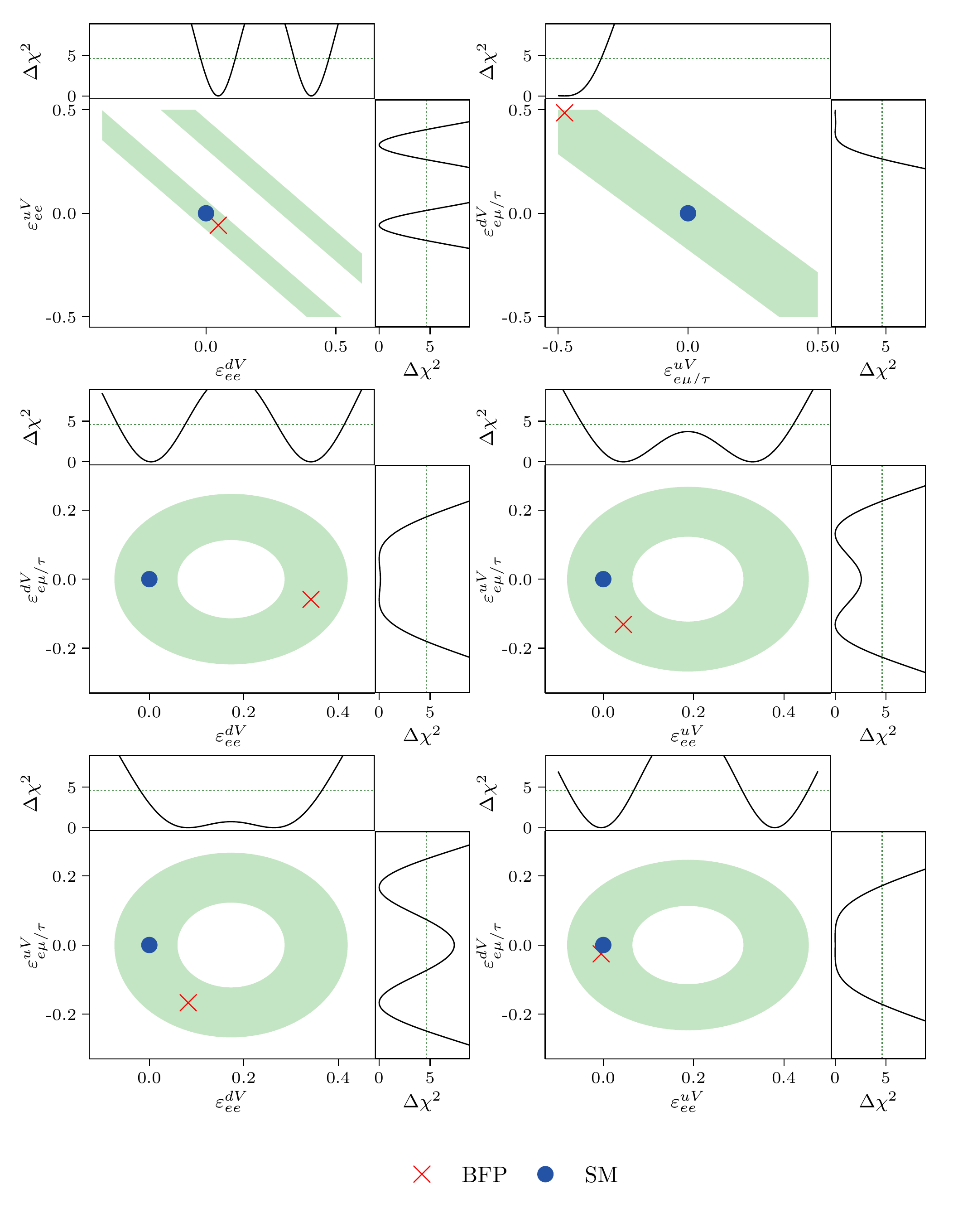}%
      }{%
        \fbox{\parbox[c][5cm][c]{0.85\linewidth}{\centering Figura pendiente:\\ \texttt{\detokenize{All_Plot_x2_NSI_Region.pdf}}}}%
      }
      \caption{\label{fig:nsi-panel}Two-dimensional confidence regions at $90\%$ C.L. for all pairings of the vector NSI parameters $\varepsilon^{qV}_{\alpha\beta}$, obtained from the profiled $\Delta\chi^2$ with two degrees of freedom. The auxiliary panels show one-dimensional $\Delta\chi^2$ cuts through the best-fit point (BFP) along the corresponding parameter axes. The overlap with the nuclear-only regions is quantified in the text.  The SM prediction, $\varepsilon^{qV}_{\alpha\beta}=0$, lies at the origin of each panel.}
    \end{figure}

    We vary each NSI coefficient separately and then study all six two-parameter combinations, profiling the normalization in each case. The resulting one-dimensional profiles and joint regions are shown in \autoref{fig:nsi-one-parameter} and \autoref{fig:nsi-panel}.

    In \autoref{fig:nsi-one-parameter} , the diagonal parameters $\varepsilon^{uV}_{ee}$ and $\varepsilon^{dV}_{ee}$ exhibit a qualitatively distinct behavior: their $\Delta\chi^2$ profiles display two separate local minima, one in the vicinity of the SM limit ($\varepsilon^{qV}_{ee}\approx 0$) and a second degenerate solution near   $\varepsilon^{qV}_{ee}\sim0.35$ --  $0.38$ . This double-minimum structure is a direct consequence of the quadratic character of $\left(Q_V^\text{NSI}\right)^2$ (see   \autoref{sec:nsi} ): the diagonal NSI parameters enter the weak charge through a linear shift of the SM vector couplings, so the squared amplitude admits two distinct parameter values that yield the same predicted event rate. At an intermediate value between the two minima, there exists a local maximum, which acts as a barrier in the $\Delta\chi^2$ landscape and produces a disconnected confidence region. 

      Flavor-changing parameters enter the nuclear rate through squared amplitudes without  interference with the SM   final flavor. Their profiles are therefore even functions of the corresponding coupling. The shallow symmetric minima remain within a single connected $1\sigma$ interval containing zero; the data do not require flavor-changing interactions.

    Relative to the nuclear-only calculation, E$\nu e$S changes the total $1\sigma$ interval length by $+0.0107\%$, $+0.0108\%$, $-0.0167\%$, and $-0.0166\%$ for $\varepsilon^{uV}_{ee}$, $\varepsilon^{dV}_{ee}$, $\varepsilon^{uV}_{e\mu/\tau}$, and $\varepsilon^{dV}_{e\mu/\tau}$, respectively. The largest shift of a $1\sigma$ endpoint is $2.25\times10^{-5}$, and the number of interval components is unchanged. Changes of the selected best-fit branch are not interpreted as physical displacements of the allowed set .

    The numerical $1\sigma$ confidence intervals extracted from the one-dimensional profiles are summarized in   \autoref{tab:nsi-comparison-conus-plus} . We compare our results with the independent \textsc{Conus}$+$ analysis of Ref.~\cite{DeRomeri:2025csu}. For completeness, we also include the COHERENT CsI+LAr bounds quoted in Ref.~\cite{DeRomeri:2022twg}, which provide a direct CE$\nu \mathcal{N}$S benchmark based on a different neutrino source and target composition.

    The corresponding two-dimensional projections are collected in   \autoref{tab:nsi-2d-projections}  and discussed below.

    \begin{table*}[htbp]
      \centering

       \caption{ One-dimensional projections of the $90\%$ C.L. two-parameter NSI  regions   fitted to  the \textsc{Conus}$+$   excess spectrum .   Values are extrema of allowed points on the combined-signal $200\times200$ grids. Open  bands  reaching the scan edges  are reported as   unresolved within the scanned domain . The intervals are projections of the corresponding two-dimensional regions and should not be interpreted as independent one-parameter fits.}
       \label{tab:nsi-2d-projections}

       \small

       \begin{tabular}{lccc}
        \toprule \toprule
        Two-parameter plane
        & First-axis projection
        & Second-axis projection
        & Topology \\
        \midrule

        $(\varepsilon^{dV}_{ee},\,\varepsilon^{dV}_{e\mu/\tau})$
        &   $\varepsilon^{dV}_{ee}\in[-0.0779,\,0.4251]$
         &   $|\varepsilon^{dV}_{e\mu/\tau}|\lesssim0.2518$
         & Closed \\

        $(\varepsilon^{dV}_{ee},\,\varepsilon^{uV}_{ee})$
        &   Scan-limited
         &   Scan-limited
         & Open degeneracy band \\

        $(\varepsilon^{dV}_{ee},\,\varepsilon^{uV}_{e\mu/\tau})$
        &   $\varepsilon^{dV}_{ee}\in[-0.0779,\,0.4251]$
         &   $|\varepsilon^{uV}_{e\mu/\tau}|\lesssim0.2729$
         & Closed \\

        $(\varepsilon^{uV}_{ee},\,\varepsilon^{dV}_{e\mu/\tau})$
        &   $\varepsilon^{uV}_{ee}\in[-0.0856,\,0.4606]$
         &   $|\varepsilon^{dV}_{e\mu/\tau}|\lesssim0.2518$
         & Closed \\

        $(\varepsilon^{uV}_{ee},\,\varepsilon^{uV}_{e\mu/\tau})$
        &   $\varepsilon^{uV}_{ee}\in[-0.0856,\,0.4606]$
         &   $|\varepsilon^{uV}_{e\mu/\tau}|\lesssim0.2729$
         & Closed \\

        $(\varepsilon^{uV}_{e\mu/\tau},\,\varepsilon^{dV}_{e\mu/\tau})$
        &   Scan-limited
         &   Scan-limited
         & Open degeneracy band \\

        \bottomrule \bottomrule
      \end{tabular}

     \end{table*}

    The agreement between  our intervals and the published \textsc{Conus}$+$ intervals in \autoref{tab:nsi-comparison-conus-plus} supports  the  present numerical implementation against the independent \textsc{Conus}$+$ analysis of Ref.~\cite{DeRomeri:2025csu}.   Residual differences may reflect  implementation choices, including the  weak nuclear form factor, here approximated as   $F_W\simeq1$, and the integration and interpolation strategies; their individual contributions have not been isolated . In both analyses, the SM point $\varepsilon^{qV}_{\alpha\beta}=0$ lies within the $1\sigma$ confidence regions.

    The comparison with COHERENT should be understood as a direct CE$\nu \mathcal{N}$S comparison, but not as a one-to-one replication of the same experimental conditions. \textsc{Conus}$+$ uses reactor $\bar{\nu}_e$ and a germanium target, whereas COHERENT uses stopped-pion neutrinos and CsI+LAr targets. The different proton-to-neutron ratios modify the degeneracy directions in the NSI parameter space, while the different flavor composition explains why COHERENT distinguishes $\varepsilon^{qV}_{e\mu}$ from $\varepsilon^{qV}_{e\tau}$. Numerically, \textsc{Conus}$+$ provides constraints comparable to COHERENT in the flavor-preserving sector and is competitive in the $e\tau$ flavor-changing sector, while COHERENT remains stronger for the $e\mu$ flavor-changing parameters due to its sizeable $\nu_\mu$ component. \cite{DeRomeri:2022twg}

    Although the one-parameter analysis provides useful bounds for individual NSI couplings, it does not capture the correlations among them. Thus, we extend the analysis to two-parameter fits, varying pairs of NSI coefficients simultaneously while profiling over the nuisance parameter $\alpha$.

     The two-dimensional regions in   \autoref{fig:nsi-panel}  complement the one-dimensional profiles and make the structure of the NSI degeneracies explicit. In the flavor-preserving plane $(\varepsilon^{uV}_{ee},\varepsilon^{dV}_{ee})$, the allowed region is organized into two approximately parallel bands. These bands are the two-dimensional manifestation of the two one-dimensional minima discussed above: one branch is connected with the SM solution, while the second branch corresponds to the non-SM solution generated by the sign degeneracy of the effective weak charge.

     In the flavor-changing plane $(\varepsilon^{uV}_{e\mu/\tau},\varepsilon^{dV}_{e\mu/\tau})$, the allowed region forms a single open band. This follows directly from the fact that flavor-changing NSI enter the CE$\nu \mathcal{N}$S rate only through the squared linear combination

    \begin{equation}
    \resizebox{0.42\textwidth}{!}{$
      \left|
      Z\left(2\varepsilon^{uV}_{e\mu/\tau}+\varepsilon^{dV}_{e\mu/\tau}\right)
      +N\left(\varepsilon^{uV}_{e\mu/\tau}+2\varepsilon^{dV}_{e\mu/\tau}\right)
      \right|^2 .
    $}
      \label{eq:nsi-flavor-changing-combination}
    \end{equation} 

    The two planes $(\varepsilon^{uV}_{ee},\varepsilon^{dV}_{ee})$ and $(\varepsilon^{uV}_{e\mu/\tau},\varepsilon^{dV}_{e\mu/\tau})$ exhibit open degeneracy bands because the CE$\nu \mathcal{N}$S rate constrains only specific linear combinations of the corresponding NSI couplings.   Within the scanned domains, these bands do not yield resolved finite individual projections . By contrast, the four mixed parameter planes display closed confidence regions, where the interplay between the flavor-preserving and flavor-changing contributions confines the allowed parameter space around the best-fit solutions. In every panel, the SM point lies inside the allowed region, showing that the \textsc{Conus}$+$ data remain fully consistent with the absence of NSI.

    At the contour level, the topology of the two-dimensional regions is consistent with the independent \textsc{Conus}$+$ analysis of Ref.~\cite{DeRomeri:2025csu} and with the degeneracy patterns discussed in earlier CE$\nu \mathcal{N}$S studies~\cite{Giunti:2019xpr,Canas:2019fjw,DeRomeri:2022twg}. The most relevant comparison with COHERENT is not a direct comparison of one-dimensional projections, but rather the complementarity of the two-dimensional bands. Because germanium, CsI, and argon have different proton-to-neutron ratios, the allowed bands in the $(\varepsilon^{dV}_{ee},\varepsilon^{uV}_{ee})$ and $(\varepsilon^{dV}_{e\mu/\tau},\varepsilon^{uV}_{e\mu/\tau})$ planes have different slopes. This complementarity is essential for breaking CE$\nu \mathcal{N}$S degeneracies in combined multi-target analyses.

    The   projections in \autoref{tab:nsi-2d-projections} describe the extent of the joint 90\% C.L. region , not independent one-parameter   confidence intervals. The two open bands reach the boundaries of the chosen domains; the grid alone does not establish their behavior at arbitrarily large couplings. For the four  closed mixed regions, the   sampled projections are approximately $[-0.0779,0.4251]$ for $\varepsilon^{dV}_{ee}$ and $[-0.0856,0.4606]$ for $\varepsilon^{uV}_{ee}$, with $|\varepsilon^{dV}_{e\mu/\tau}|\lesssim0.2518$ and $|\varepsilon^{uV}_{e\mu/\tau}|\lesssim0.2729$ .

      Using \eqref{eq:region-comparison}, the six nuclear-only and combined regions have Jaccard overlap between $99.9707\%$ and $100\%$, and their allowed areas differ by at most $0.0293\%$. The sampled projections are identical between channels. These comparisons are restricted to the common scanned domains, particularly for the open bands, and establish negligible changes at the evaluated resolution rather than exact equality of continuous contours.

 \subsection{Light vector \texorpdfstring{$Z'$}{Z'} mediators}
 \label{sec:vector-results}
  We next constrain the mediator mass and coupling using the combined nuclear and electronic prediction. The angular  $E_6$   scan tests charge dependence at fixed coupling, whereas the benchmark mass--coupling scans resolve the different momentum-transfer scales of the two scattering channels.

\subsubsection{\texorpdfstring{$E_6$}{E6} Models}
\label{sec:e6-results}
    Grand-unified extensions based on groups such as $E_6$ predict additional neutral gauge bosons ($Z'$). In the two-angle $E_6$ basis, the model dependence is encoded in the angular parameters $(\alpha,\beta)$, which select the benchmark direction in the $E_6$ parameter space. For the numerical analysis, we adopt the conventional unification-motivated normalization $g_{Z'}=0.46$, following the normalization used in Ref.~\cite{Langacker:2008yv}.
  
  For comparison, we also consider $g_{Z'}=0.46$, corresponding to the $E_6$ benchmark commonly used in collider analyses. We emphasize that this benchmark is already subject to important constraints from collider searches, \cite{ATLAS:2019erb} particularly through the couplings of the $Z'$ to quarks and charged leptons. Our purpose in displaying this benchmark is therefore not to claim that \textsc{Conus}$+$ supersedes existing collider limits, but rather to illustrate the complementary information provided by coherent elastic neutrino-nucleus scattering. In contrast to conventional resonance searches at the LHC and LEP, \textsc{Conus}$+$ is directly sensitive to the $Z'\nu\nu$ interaction through neutrino scattering on nuclei. Consequently, it probes combinations of $Z'$ couplings involving neutrinos that are accessed differently, or only indirectly, in charged-lepton and hadronic collider searches. Moreover, for the smaller couplings considered in our analysis, the $Z'$ width scales as $\Gamma_{Z'}\propto g_{Z'}^2$ and can become extremely narrow, reaching values of $\Gamma_{Z'}/M_{Z'}$ of order $10^{-6}$. Our results are therefore complementary to existing collider constraints, particularly in the small-coupling regime and in the neutrino-interaction sector.
  
\begin{figure}[htbp]
    \centering
    \includegraphics[width=\linewidth]{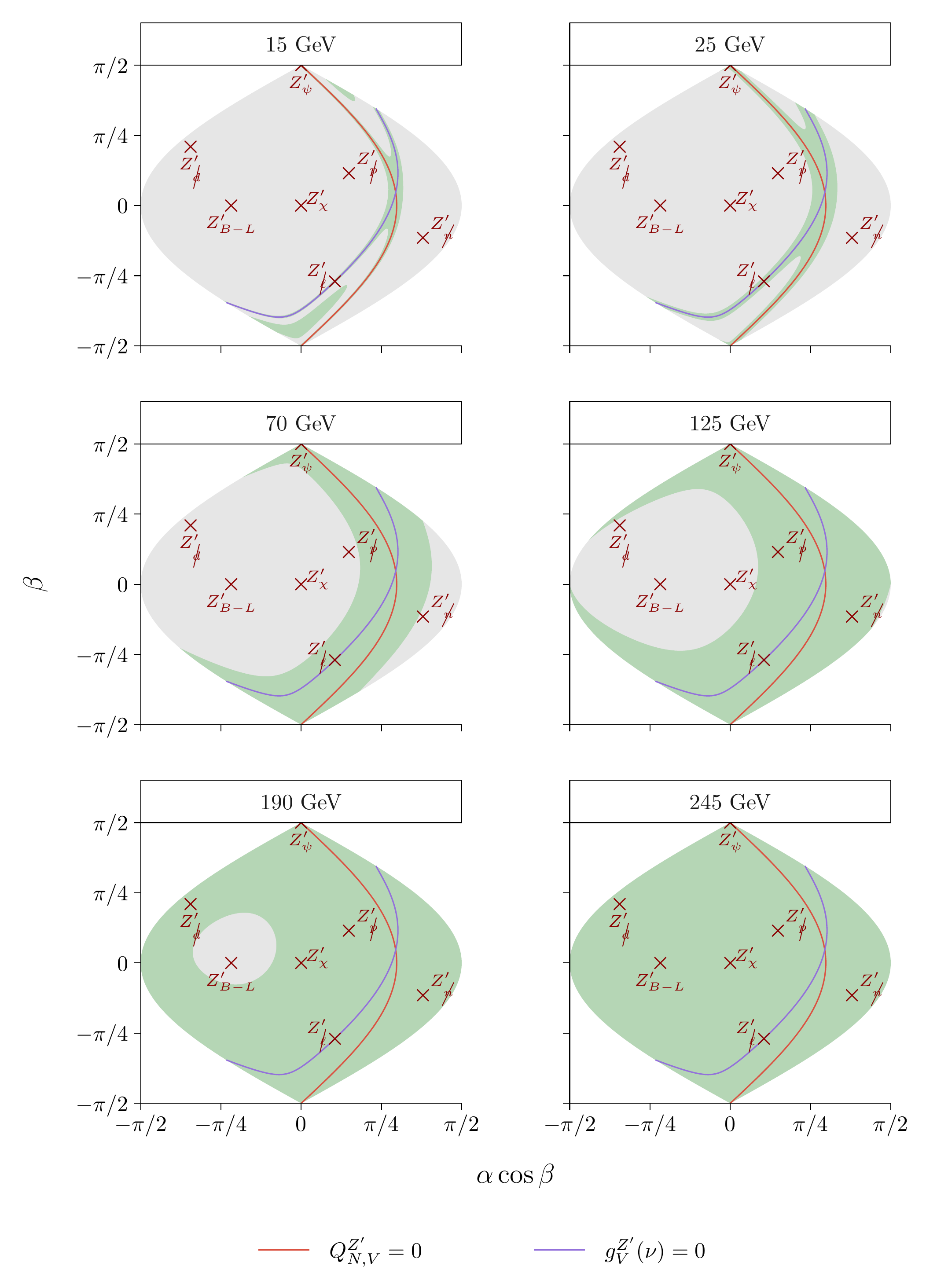}
    \caption{\label{fig:e6-regions-collage}Allowed regions in the Sanson--Flamsteed projection of the $E_6$ angular parameter space for six representative $Z'$ masses, obtained with the canonical normalization $g_{Z'}=\sqrt{5/3}\,g\tan\theta_W\simeq 0.46$~\cite{Langacker:2008yv}. Each panel displays the configurations compatible with the \textsc{Conus}$+$ data at $90\%$ confidence level for two degrees of freedom. The horizontal and vertical coordinates are $\alpha\cos\beta$ and $\beta$, respectively, with $\alpha,\,\beta\in[-\pi/2,\,\pi/2]$. The constraints are strongest for the lightest mediator masses and become progressively weaker as the mass increases, until the scanned region is nearly unconstrained for the largest values of $m_{Z'}$. Benchmark cases are listed in Table~\ref{tab:e6-model-charges}.}
  \end{figure}

  The allowed regions in the $(\alpha,\beta)$ plane for six representative $Z'$ masses are shown in Fig.~\ref{fig:e6-regions-collage}. The main trend is clear: for light mediators, the excluded part of the parameter space is larger, while the allowed region becomes progressively broader as the $Z'$ mass increases. This behavior follows from the propagator suppression of the new contribution to CE$\nu \mathcal{N}$S: when $m_{Z'}$ becomes large compared with the typical momentum transfer, the effect of the new boson on the recoil spectrum decreases. In the largest-mass panels, the contours nearly saturate the scanned domain, indicating that the present \textsc{Conus}$+$ dataset loses sensitivity to the angular structure of the $E_6$ benchmarks in that regime.
  
  The benchmark directions listed in Table~\ref{tab:e6-model-charges} provide a useful interpretation of this pattern. In general, configurations that suppress either the effective neutrino coupling or the coherent nuclear vector charge are less efficiently constrained, whereas benchmark points with stronger effective couplings are more readily excluded by the data.
  
  For the angular scan, adding the elastic antineutrino--electron scattering contribution, E$\nu e$S, produces nearly coincident allowed regions at the evaluated resolution. Across the six masses considered, the change in the projected allowed area is at most $0.0214\%$ and the Jaccard overlap between the nuclear-only and combined regions is at least $99.9129\%$. Thus, the angular interpretation remains practically unchanged. This conclusion applies to the fixed-coupling scan at GeV mediator masses and does not imply that E$\nu e$S is negligible in the light-mediator mass--coupling analysis discussed below.
  
  A complementary view is given in Fig.~\ref{fig:light-vector-benchmarks}, where the results are displayed directly in the $(m_{Z'},g_{Z'})$ plane. Panel~(a) of Fig.~\ref{fig:light-vector-benchmarks} shows the exclusion envelopes for the benchmark $E_6$ models of Table~\ref{tab:e6-model-charges}, while panel~(b) compares the combined constraints for the leptophilic $L_e-L_\mu$ and $L_e-L_\tau$ models with their common electronic reference. In both cases, the exclusion contours display the characteristic behavior of a light vector mediator. In the low-mass regime, $m_{Z'}^2\ll |q^2|$, the propagator is dominated by the momentum transfer, and the dependence on the mediator mass becomes weak. As a result, the exclusion boundary approaches an approximately constant value, producing a horizontal low-mass plateau. Conversely, for $m_{Z'}^2\gg |q^2|$, the propagator suppresses the new-physics contribution, and increasingly larger values of $g_{Z'}$ are required to generate an observable distortion of the recoil spectrum. The loss of sensitivity at large mediator masses is therefore a common feature of both the $E_6$ and leptophilic scenarios.
  
  For the $E_6$ benchmarks, the separation among the exclusion envelopes is mainly driven by the different charge assignments. The relevant scaling is controlled by the effective product $g^{Z'}_V(\nu)Q^{Z'}_{N,V}$ entering the CE$\nu \mathcal{N}$S amplitude. Benchmarks with larger effective products generate larger spectral distortions for the same value of $g_{Z'}$ and are therefore constrained at smaller couplings. Conversely, benchmarks with suppressed effective charges or partial cancellations require larger couplings to reach the same statistical exclusion threshold.
  
  The corresponding low-mass plateau values, extracted from the nuclear-only numerical scan, are summarized in Table~\ref{tab:e6-nuclear-plateau-limits}. The quantity $g_{Z'}^{\rm lim}$ represents the approximately constant upper limit in the light-mediator regime, while the quoted mass range indicates the interval over which the exclusion boundary remains compatible with the initial flat segment.
  \begin{table}[htbp]
    \centering
    \caption{\label{tab:e6-nuclear-plateau-limits}Low-mass plateau of the lower, first-crossing nuclear-only $90\%$ C.L. exclusion boundaries for the benchmark $E_6$ models in the $(m_{Z'},g_{Z'})$  considering CE$\nu \mathcal{N}$S only. The quantity $g_{Z'}^{\rm lim}$ denotes the approximately constant upper limit in the regime $m_{Z'}^2\ll |q^2|$, where the sensitivity is nearly independent of the mediator mass. The third column gives the approximate upper bound of this regime, corresponding to the point at which $m_{Z'}$ becomes comparable to the characteristic momentum-transfer scale probed by \textsc{Conus}$+$ and the exclusion contour begins to rise.}
    \begin{tabular}{ccc}
      \toprule \toprule
      Benchmark model & $g_{Z'}^{\rm lim}$ & Low-mass plateau range \\
      \midrule
      $Z'_{B-L}$ & $2.703\times 10^{-5}$ & $m_{Z'}\lesssim 3.571\,\mathrm{MeV}$ \\
      $Z'_{\chi}$ & $3.425\times 10^{-5}$ & $m_{Z'}\lesssim 3.571\,\mathrm{MeV}$ \\
      $Z'_{\cancel{d}}$ & $3.905\times 10^{-5}$ & $m_{Z'}\lesssim 3.571\,\mathrm{MeV}$ \\
      $Z'_{\cancel{n}}$ & $7.057\times 10^{-5}$ & $m_{Z'}\lesssim 3.571\,\mathrm{MeV}$ \\
      $Z'_{\cancel{p}}$ & $6.258\times 10^{-5}$ & $m_{Z'}\lesssim 3.571\,\mathrm{MeV}$ \\
      \bottomrule \bottomrule
    \end{tabular}
  \end{table}
  
  The complete prediction also includes E$\nu e$S. Since the electronic and nuclear channels probe the mediator propagator through the different momentum-transfer scales $m_{Z'}^2+2m_eT_e$ and $m_{Z'}^2+2m_{\mathcal N}T_{\mathcal N}$, respectively, the combined exclusion boundary exhibits two distinct regimes. At sufficiently low mass, E$\nu e$S determines a stronger, lower plateau. As the mediator mass increases, the electronic contribution loses sensitivity first and the boundary approaches a higher, approximately horizontal nuclear-dominated shoulder. At still larger masses, both channels are propagator suppressed and the coupling limit rises.
  
  The combined low-mass plateaus and the representative limits at $m_{Z'}=3\,\mathrm{MeV}$ are given in Table~\ref{tab:e6-plateau-limits}. The latter mass lies within the nuclear-dominated shoulder of all five benchmarks. The shoulder ranges identify the sampled intervals in which the combined limit is within $5\%$ of both the nuclear-only plateau and the nuclear-only limit at the same mass; they characterize the boundary shape and are not additional confidence intervals.
  \begin{table*}[htbp]
    \centering
    \caption{\label{tab:e6-plateau-limits}Characteristic combined coupling limits at $90\%$ C.L. for the $E_6$ benchmarks. The second column is the low-mass plateau dominated by E$\nu e$S, whereas the third column gives the combined limit at $m_{Z'}=3\,\mathrm{MeV}$, representative of the nuclear-dominated shoulder.}
    \begin{tabular}{lccc}
      \toprule \toprule
      Benchmark & Combined low-mass plateau & Combined limit at $3\,\mathrm{MeV}$ & Shoulder range [MeV] \\
      \midrule
      $Z'_{B-L}$ & $7.487\times10^{-7}$ & $2.796\times10^{-5}$ & $[0.901,3.571]$ \\
      $Z'_{\chi}$ & $1.116\times10^{-6}$ & $3.539\times10^{-5}$ & $[0.766,3.571]$ \\
      $Z'_{\cancel d}$ & $9.163\times10^{-7}$ & $4.032\times10^{-5}$ & $[1.060,3.571]$ \\
      $Z'_{\cancel n}$ & $1.263\times10^{-6}$ & $7.288\times10^{-5}$ & $[1.351,3.571]$ \\
      $Z'_{\cancel p}$ & $1.594\times10^{-6}$ & $6.471\times10^{-5}$ & $[0.977,3.571]$ \\
      \bottomrule \bottomrule
    \end{tabular}
  \end{table*}
  
  The strongest low-mass constraint is obtained for $Z'_{B-L}$, with $g_{Z'}^{\rm lim}=2.703\times10^{-5}$ for $m_{Z'}\lesssim3.571\,\mathrm{MeV}$ in the nuclear-only calculation. This result is consistent with the evolution of CE$\nu \mathcal{N}$S bounds on $U(1)_{B-L}$ vector mediators reported in the literature. Earlier analyses based on COHERENT data typically reached sensitivities of order $10^{-4}$--$10^{-3}$ for MeV-scale mediators~\cite{Miranda:2020tif,Cadeddu:2020nbr}, whereas reactor-based measurements with CONUS and the first \textsc{Conus}$+$ result have pushed the sensitivity into the $10^{-5}$ regime~\cite{DeRomeri:2025csu,CONUS:2026uhz}. The present $B-L$ plateau therefore lies within the expected range of modern low-threshold reactor CE$\nu \mathcal{N}$S constraints.
  
  It is followed by $Z'_{\chi}$ and $Z'_{\cancel{d}}$, whose plateau limits are $3.425\times10^{-5}$ and $3.905\times10^{-5}$, respectively, both for $m_{Z'}\lesssim3.571\,\mathrm{MeV}$. The weakest low-mass bounds correspond to $Z'_{\cancel{p}}$ and $Z'_{\cancel{n}}$, with plateau values around $6\times10^{-5}$. This ordering should not be interpreted as a statistical preference for one benchmark over another; rather, it is a consequence of the different combinations of $g^{Z'}_V(\nu)$ and $Q^{Z'}_{N,V}$ entering the CE$\nu \mathcal{N}$S amplitude. Unlike the $B-L$ case, dedicated CE$\nu \mathcal{N}$S limits for the other benchmark directions are not commonly available in the literature. The present analysis therefore extends the \textsc{Conus}$+$ sensitivity study to a broader class of $E_6$-motivated scenarios.
  
  When E$\nu e$S is included, the strongest low-mass bound remains that of $Z'_{B-L}$, but the ordering of the other benchmark limits changes because the electron and nuclear charges differ. The electronic contribution strengthens the low-mass coupling limits by factors of approximately $31$--$56$ relative to the nuclear-only plateaus. The combined low-mass plateau extends from the scan minimum $10^{-3}\,\mathrm{MeV}$ to approximately $5.05\times10^{-3}\,\mathrm{MeV}$, while at $m_{Z'}=3\,\mathrm{MeV}$ the combined boundaries agree with the nuclear-only results to within $0.29\%$, confirming the nuclear origin of the higher shoulder.
  
  For mediator masses above the corresponding plateau range, the exclusion envelopes begin to rise. At the upper edge of the scan, $m_{Z'}=10^4\,\mathrm{MeV}$, the exclusion boundary has already moved to much larger couplings, reaching values between $g_{Z'}\simeq2.331\times10^{-2}$ for $Z'_{B-L}$ and $g_{Z'}\simeq6.083\times10^{-2}$ for $Z'_{\cancel{n}}$. Panel~(a) of Fig.~\ref{fig:light-vector-benchmarks} should therefore be read as a benchmark-dependent translation of the \textsc{Conus}$+$ spectral information into exclusion limits on $(m_{Z'},g_{Z'})$, rather than as evidence for a preferred $E_6$ realization.
  
  For the combined prediction, the corresponding limits at $m_{Z'}=10^4\,\mathrm{MeV}$ range from $2.331\times10^{-2}$ for $Z'_{B-L}$ to $6.083\times10^{-2}$ for $Z'_{\cancel n}$. Thus, the pronounced improvement at very low mass cannot be inferred from the small changes in the best-fit pulls or in the fixed-coupling angular regions.

\subsubsection{Leptophilic models}
\label{sec:lalb-results}
  \begin{figure*}[htbp]
     \centering
     \subfloat[$E_6$ benchmark boundaries.\label{fig:e6-models-sub}]{%
      \IfFileExists{Pictures/E6_Models/Plot_E6_CEvNS_EveS.pdf}{%
        \includegraphics[width=0.5\linewidth]{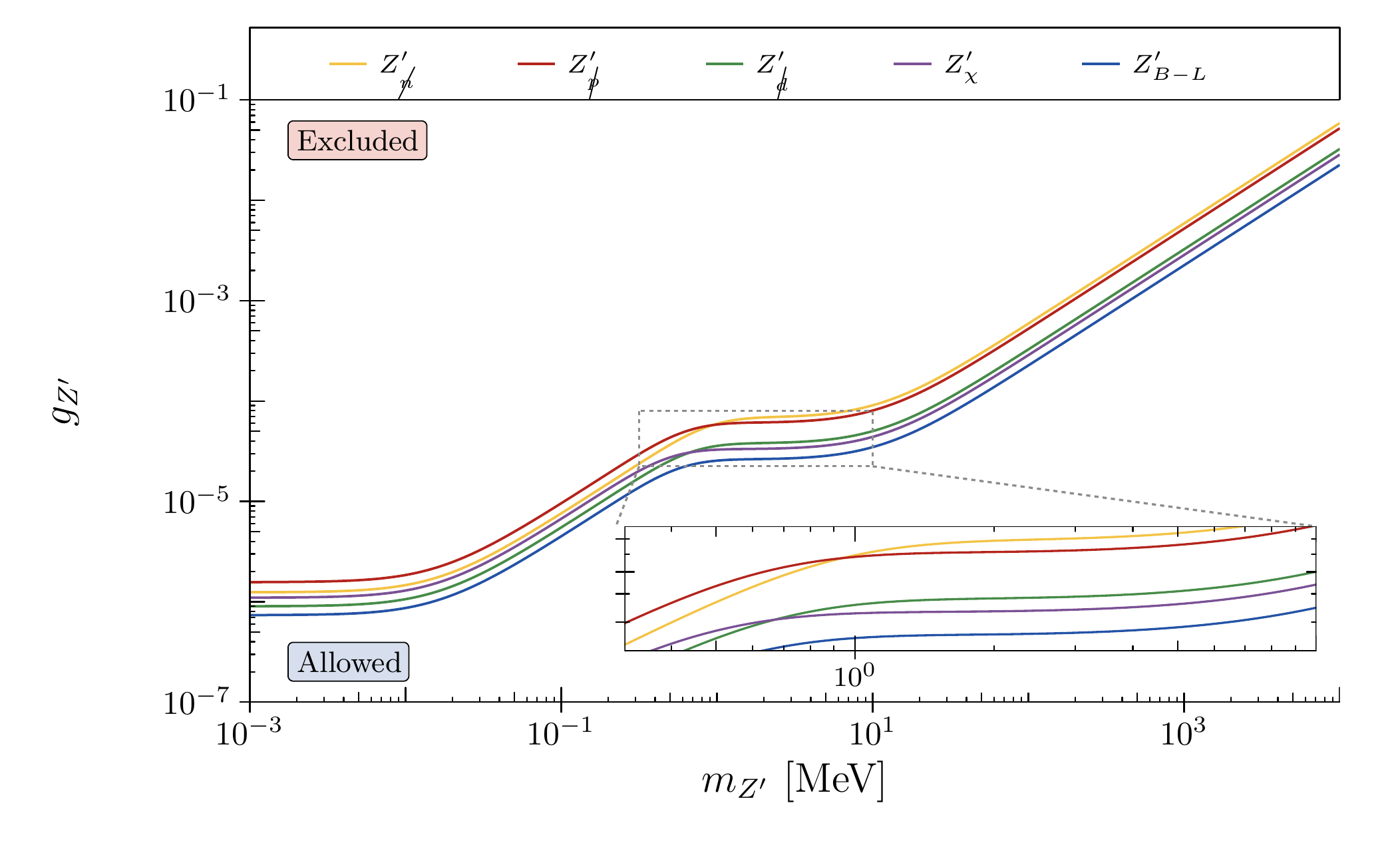}%
      }{%
        \fbox{\parbox[c][5cm][c]{0.46\linewidth}{\centering Figura pendiente:\\ \texttt{\detokenize{Plot_E6_CEvNS_EveS.pdf}}}}%
      }}
    \hfill
    \subfloat[Leptophilic boundaries.\label{fig:leptophilic-region-sub}]{%
      \IfFileExists{Pictures/LaLb/Plot_LaLb_CEvNS_EveS.pdf}{%
        \includegraphics[width=0.5\linewidth]{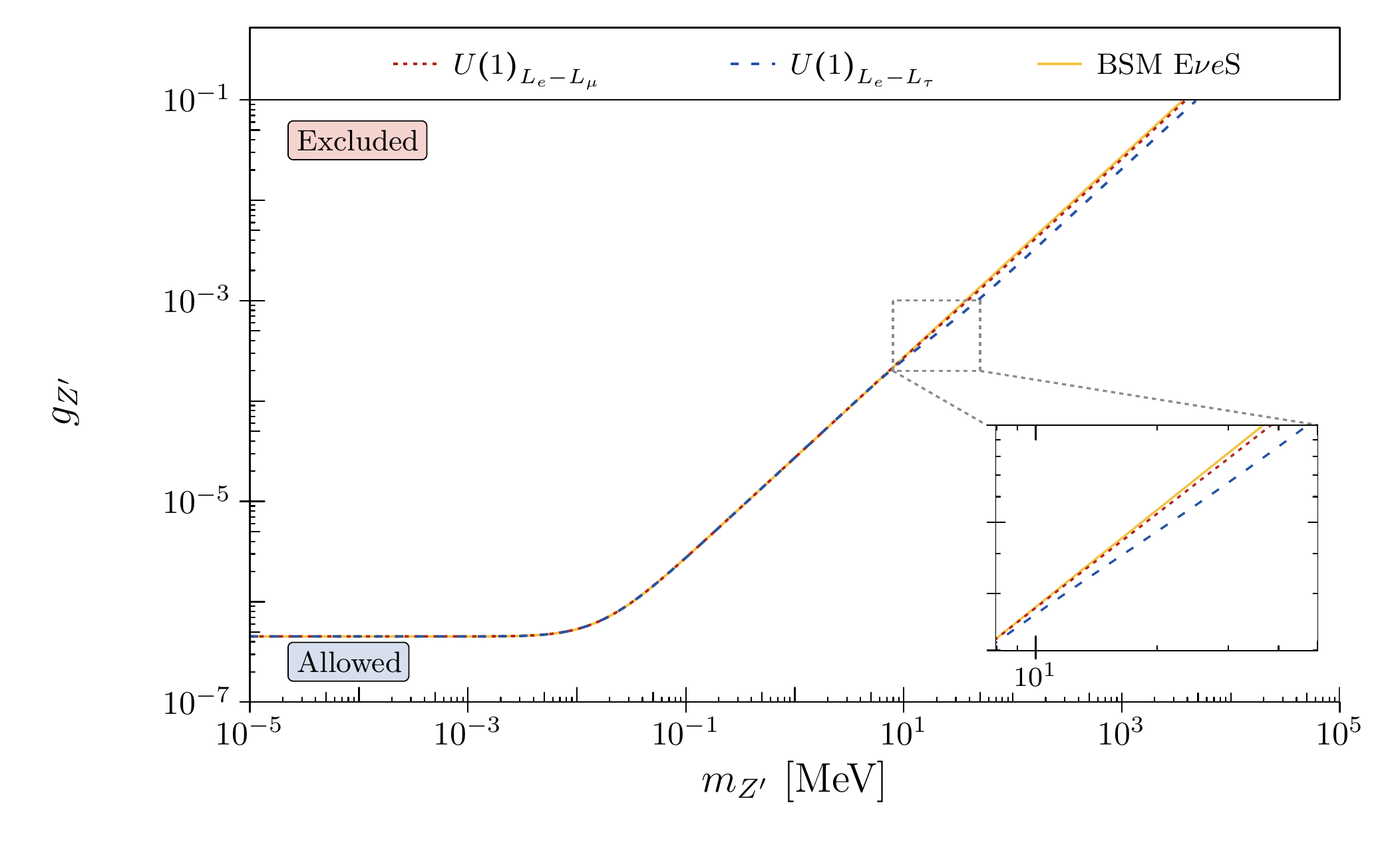}%
      }{%
        \fbox{\parbox[c][5cm][c]{0.46\linewidth}{\centering Figura pendiente:\\ \texttt{\detokenize{Plot_LaLb_CEvNS_EveS.pdf}}}}%
      }}
     \caption{  Mediator mass--coupling boundaries at 90\%  C.L.   Panel (a) shows  the   five  $E_6$   benchmarks using  the   combined CE$\nu\mathcal N$S+E$\nu e$S prediction .   Panel (b) compares  the   complete $L_e-L_\mu$ and $L_e-L_\tau$ predictions with their common electronic reference. The complete leptophilic boundaries contain SM and $Z'$ effects  in   both scattering channels ,   whereas  the   reference retains both SM channels but allows  the   $Z'$ to modify only E$\nu e$S . The  upper   side  of   each boundary is excluded within  the   scanned domain. Insets enlarge  the   regions indicated  by  the   gray rectangles and their connectors .}
      \label{fig:light-vector-benchmarks}
  \end{figure*}

  \begin{table*}[htbp]
    \centering
    \caption{Coupling limits at 90\% C.L. for the electronic reference and the complete leptophilic predictions at representative masses. The electronic reference retains both SM channels but allows the $Z'$ to modify only E$\nu e$S. For each complete model, the improvement is $100(1-g_{\mathrm{full}}^{\mathrm{lim}}/g_{\mathrm{E}\nu e\mathrm{S}}^{\mathrm{lim}})$; a positive value indicates that the loop-induced nuclear mediator contribution strengthens the bound, whereas a negative value indicates a weakening. Values are interpolated in logarithmic mass and coupling from the unsmoothed boundaries.}
    \label{tab:lalb-limits}
    \begin{tabular}{cccccc}
       \toprule \toprule 
        $m_{Z'}$ [MeV]
       &   E$\nu e$S reference
       &   $L_e-L_\mu$
      & Improvement$_\mu$ [\%]
      & $L_e-L_\tau$
      & Improvement$_\tau$ [\%]  \\
      \midrule
        1  &   $2.7588\times10^{-5}$  &   $2.7591\times10^{-5}$  &   $-0.010$  &   $2.7594\times10^{-5}$ & $-0.020$  \\
        10  &   $2.7580\times10^{-4}$  &   $2.7284\times10^{-4}$ & $+1.072$ & $2.6191\times10^{-4}$ & $+5.036$  \\
        100  &   $2.7561\times10^{-3}$  &   $2.5670\times10^{-3}$ & $+6.860$ & $2.0239\times10^{-3}$ & $+26.567$  \\
        1000  &   $2.7595\times10^{-2}$  &   $2.5620\times10^{-2}$ & $+7.156$ & $2.0100\times10^{-2}$ & $+27.162$  \\
      \bottomrule \bottomrule
    \end{tabular}
    \end{table*}
  The $L_e-L_\mu$ and $L_e-L_\tau$ models couple directly to electrons and reactor antineutrinos, whereas their nuclear coupling arises through the loop-induced photon--$Z'$ mixing in \autoref{sec:leptophilic-mixing}. We include both contributions in panel (b) of \autoref{fig:light-vector-benchmarks} to determine when the nuclear response distinguishes models whose direct electronic signal is identical. Both channels retain their SM and mediator contributions, including interference.

  The   combined  low-mass   plateaus are indistinguishable at the quoted precision, with $g_{Z'}^{\mathrm{lim}}=4.586\times10^{-7}$ for both models. Under the 5\% criterion, this regime extends from the scan minimum  $10^{-5}$   MeV to approximately $4.61\times10^{-3}$ MeV. The direct electronic contribution dominates here, explaining the coincidence without requiring the nuclear loop amplitudes to be identical. The subsequent separation and the effect of the nuclear mediator contribution are quantified relative to the electronic reference in \autoref{tab:lalb-limits}.

    To isolate the nuclear mediator contribution, we also evaluate a reference prediction containing both SM channels but allowing the $Z'$ to modify only E$\nu e$S. This reference yields the same electronic boundary  for   $L_e-L_\mu$ and $L_e-L_\tau$. Adopting a $1\%$ relative change in the coupling limit as a practical criterion, the loop-induced nuclear contribution is negligible for both models up to approximately $m_{Z'}=5.9$ MeV. The individual $1\%$ thresholds are approximately $9.8$ MeV for $L_e-L_\mu$  and   $5.9$ MeV for $L_e-L_\tau$. At $m_{Z'}=100$ MeV, by contrast, the electronic reference limit is $2.7561\times10^{-3}$, and including the nuclear mediator contribution lowers it by $6.86\%$  for   $L_e-L_\mu$  and   $26.57\%$  for   $L_e-L_\tau$ .  Consequently, the   loop suppression does not make the nuclear response irrelevant throughout the scan. Its flavor-dependent charged-lepton masses account for the separation of the combined boundaries, while interference and the independent normalization profiles determine the direction and size of the change .

  The   comparison in \autoref{tab:lalb-limits} is restricted to masses with resolved crossings of  the   adopted threshold. Where a boundary leaves the scanned coupling domain, we do not extrapolate an exclusion limit. Moreover, a nuclear-only leptophilic scan can contain disconnected allowed branches, so its upper envelope should not  be interpreted as  a single connected allowed interval. The combined boundaries shown here have a single resolved crossing at  the   masses used in the table.

\section{Conclusions}
\label{sec:conclusions}

  We have performed a binned-$\chi^2$ analysis of the first \textsc{Conus}$+$ measurement of coherent elastic antineutrino--nucleus scattering on germanium. The analysis employs a profiled nuisance-parameter treatment of the global normalization uncertainty, allowing the SM, NSI, $E_6$-motivated light-$Z'$ scenarios, and the leptophilic $U(1)_{L_e-L_\mu}$ and $U(1)_{L_e-L_\tau}$ models to be studied within a common statistical framework. In the revised prediction, the reconstructed-energy spectrum contains both coherent elastic antineutrino--nucleus scattering, CE$\nu\mathcal N$S, and elastic antineutrino--electron scattering, E$\nu e$S. The SM contribution and the mediator interference are retained in each channel, while a common normalization nuisance parameter is analytically profiled.
  
  The \textsc{Conus}$+$ data are fully consistent with the SM prediction. The measured excess is compatible with the expected CE$\nu \mathcal{N}$S signal, the pull distribution exhibits no coherent spectral distortion, and the best-fit $\chi^2$ values differ only marginally among the SM and the BSM scenarios considered. After accounting for the number of fitted parameters, the SM remains the most parsimonious description of the present \textsc{Conus}$+$ dataset. The inclusion of E$\nu e$S changes the best-fit statistics and binwise pulls only negligibly in the neighborhood of the SM solution; nevertheless, its impact on exclusion boundaries for sufficiently light mediators is substantial.
  
  In the NSI sector, the one-dimensional analysis reproduces the characteristic CE$\nu \mathcal{N}$S degeneracy pattern. The flavor-preserving parameters exhibit a two-branch structure, reflecting the quadratic dependence of the rate on the effective weak charge, whereas the flavor-changing parameters remain near the SM point. The resulting $1\sigma$ confidence intervals are summarized in Table~\ref{tab:nsi-comparison-conus-plus}, where they are shown to be in excellent agreement with the independent \textsc{Conus}$+$ analysis of Ref.~\cite{DeRomeri:2025csu}. In all cases, the SM point lies within the allowed region, indicating that the present data constrain NSI couplings but provide no evidence for non-standard neutrino interactions. For the quark NSI considered here, the electronic contribution remains purely SM; consequently, the combined treatment leaves the NSI regions practically unchanged while providing a common description for all scenarios.
  
  The two-dimensional analysis further clarifies the structure of the NSI parameter space. The flavor-preserving and flavor-changing planes exhibit open degeneracy bands, while the mixed planes yield closed confidence regions with finite projections. These projected intervals, reported in Table~\ref{tab:nsi-2d-projections}, should be interpreted as projections of two-dimensional confidence regions rather than independent one-parameter bounds. Comparison with COHERENT highlights the complementarity between reactor and stopped-pion CE$\nu \mathcal{N}$S measurements. The different proton-to-neutron ratios of Ge, CsI, and Ar produce different degeneracy directions in the $(\varepsilon^{uV},\varepsilon^{dV})$ parameter space, demonstrating that combined multi-target analyses will be essential for improving future NSI constraints.
  
  Within the $E_6$ framework, the two-angle analysis, displayed through the Sanson--Flamsteed projection, provides a geometric interpretation of the \textsc{Conus}$+$ sensitivity by identifying the regions of the $(\alpha,\beta)$ plane that are compatible with the data. Light mediators exclude sizeable portions of the parameter space, while sensitivity decreases progressively with increasing mediator mass due to propagator suppression. The corresponding exclusion limits for the benchmark $E_6$ models are summarized in Table~\ref{tab:e6-plateau-limits}. Among the scenarios considered, the $Z'_{B-L}$ direction provides the strongest constraints. For the fixed-coupling angular scans, the nuclear-only and combined CE$\nu\mathcal N$S+E$\nu e$S regions are nearly identical at the evaluated resolution. In contrast, the mass--coupling boundaries show two distinct low-mass regimes: an electronic plateau at very small masses and a higher nuclear-dominated shoulder at MeV masses. The low-mass limits are strengthened by factors of approximately $31$--$56$ with respect to the nuclear-only calculation; the strongest combined bound is $g_{Z'}^{\rm lim}=7.487\times10^{-7}$ for the $Z'_{B-L}$ benchmark.
  
  For the leptophilic $U(1)_{L_e-L_\mu}$ and $U(1)_{L_e-L_\tau}$ models, the additional $Z'$-mediated CE$\nu \mathcal{N}$S amplitude arises through loop-induced photon--$Z'$ kinetic mixing, which generates an effective coupling to the nuclear electromagnetic current. This amplitude interferes with the SM nuclear amplitude, and the constraints follow from the combined CE$\nu\mathcal N$S+E$\nu e$S prediction. Although the low-mass limits are dominated by direct E$\nu e$S, the loop-induced nuclear contribution becomes relevant at larger mediator masses and distinguishes the two leptophilic models through their different charged-lepton masses. At very low mass, the two models share an indistinguishable electronic plateau, $g_{Z'}^{\rm lim}=4.586\times10^{-7}$; at larger masses, their limits separate because the loop-induced nuclear contributions depend on different charged-lepton masses.
  
  Overall, the first \textsc{Conus}$+$ CE$\nu \mathcal{N}$S measurement shows no statistically significant preference for NSI, $E_6$-motivated vector mediators, or leptophilic light-$Z'$ scenarios. Instead, it provides competitive constraints on these BSM frameworks while remaining fully consistent with the SM. Future reactor CE$\nu \mathcal{N}$S measurements with larger exposures, lower detection thresholds, improved quenching-factor calibration, and reduced normalization uncertainties are expected to further strengthen these constraints. In addition, the complementarity observed between \textsc{Conus}$+$ and COHERENT suggests that combined analyses of reactor and stopped-pion CE$\nu \mathcal{N}$S data using multiple nuclear targets constitute a promising strategy for lifting NSI degeneracies and improving the sensitivity to light vector mediators. The present results also show that a small SM electronic signal should not be used as a criterion to neglect E$\nu e$S in light-mediator searches: its BSM contribution can control the exclusion limit in the low-mass regime.

\begin{acknowledgments}
  We thank Omar Miranda Romagnoli for helpful comments. This article is derived from a master's thesis in Applied Statistics at the Universidad de Nariño, Colombia, entitled \textit{``Estimación estadística de regiones de confianza para interacciones no estándar y bosones $Z'$ ligeros a partir de datos CE$\nu \mathcal{N}$S del experimento \textsc{Conus}$+$''}. Ch.M.B., C.S.M. and E.R. acknowledge the Universidad de Nariño for its academic support during the development of this work, as well as support from the Vicerrectoría de Investigaciones e Interacción Social (VIIS) of the Universidad de Nariño under project numbers 3595 and 3899. B.C. and E.R. acknowledge support from the Minciencias Grant CD 82315 CT ICETEX 2021-1080.
\end{acknowledgments}

\bibliographystyle{apsrev4-2}
\bibliography{References.bib}
\end{document}